\documentclass[letterpaper,9pt, onecolumn, twoside, nosectionnumbsers]{classes/LTJournalArticle}

\def\doctitle{Nanophotonic quantum magnetometry in a spin-dense diamond cavity}

\def\authorOne{Nicholas J. Sorensen}
\def\authorTwo{Elham Zohari}
\def\authorThree{Joshua S. Wildeman}
\def\authorFour{Sigurd Fl{\aa}gan}
\def\authorFive{Vinaya~K.~Kavatamane}
\def\authorSix{Paul E. Barclay}
\def\addressOne{
    Institute for Quantum Science and Technology, University of Calgary, Calgary, AB, T2N 1N4, Canada
    }
\def\addressTwo{Department of Physics, University of Alberta, Edmonton, AB, T6G 2R3, Canada}
\def\addressThree{National Research Council of Canada, Quantum and Nanotechnology Research Centre, Edmonton, AB, T6G 2M9, Canada}
\def\emailContact{nicholas.sorensen@ucalgary.ca} 

\def\abstractText{
Quantum sensors based on the nitrogen-vacancy (NV) center in diamond are leading platforms for high-sensitivity magnetometry with nanometer-scale resolution. State-of-the-art implementations, however, typically rely on bulky free-space optics or sacrifice spatial resolution to achieve high sensitivities.
Here, we realize an integrated platform that overcomes this trade-off by fabricating monolithic whispering-gallery-mode cavities from a diamond chip containing a high density of NV centers and by evanescently coupling excitation to and photoluminescence from the cavity using a tapered optical fiber. 
Employing a lock-in-amplified Ramsey magnetometry scheme, we achieve a photon-shot-noise-limited DC sensitivity of $58\,\text{nT}/\sqrt{\text{Hz}}$---the best sensitivity reported to date for a nanofabricated cavity-based magnetometer.
The microscopic cavity size enables micrometer-scale spatial resolution and low-power operation, while fiber-coupling provides a path to scalable on-chip integration.
Arrays of such sensors could enable NV-NMR spectroscopy of sub-nanoliter samples, new magnetic-gradient imaging architectures, and compact biosensing platforms.
}
\usepackage{hyperref}

\usepackage{graphicx}%
\usepackage{multirow}%
\usepackage{amsmath}%
\usepackage[dvipsnames]{xcolor}
\usepackage{textcomp}%
\usepackage{dcolumn}

\usepackage{orcidlink,breqn,enumitem,ulem,titlesec,lipsum,mathtools,gensymb,setspace, float,dblfloatfix}

\usepackage[nameinlink]{cleveref}
\usepackage{upgreek}

\usepackage[utf8]{inputenc}
\usepackage[T1]{fontenc}
\usepackage{mathptmx}
\usepackage{etoolbox}
\usepackage[dvipsnames]{xcolor}
\usepackage{soul}
\usepackage{ulem}
\usepackage{array}
\usepackage{enumitem}

\Crefname{equation}{Eq.}{Eqs.}
\Crefname{figure}{Fig.}{Figs.}
\Crefname{table}{Tab.}{Tabs.}
\Crefname{section}{Appendix}{Appendices}

\begin{document}

\title{\doctitle}

\author{\normalsize
\authorOne,\textsuperscript{1,*}\orcidlink{0000-0002-4666-5791} 
\authorTwo,\textsuperscript{1,2,3}\orcidlink{0000-0001-8344-4157} 
\authorThree,\textsuperscript{1} 
\authorFour,\textsuperscript{1}\orcidlink{0000-0003-0272-7601} \\   \normalsize
\authorFive,\textsuperscript{1}\orcidlink{0000-0002-4287-9290} and 
\authorSix\textsuperscript{1}\orcidlink{0000-0002-9659-5883}}
\date{\small\textit{
\textsuperscript{1}\addressOne  \\ 
\textsuperscript{2}\addressTwo \\ 
\textsuperscript{3}\addressThree \\ \textsuperscript{*}\emailContact}}

\twocolumn[
\maketitle
\begin{abstract}
    \abstractText
\end{abstract}
]

Quantum sensors leverage the quantum nature of light and matter to enable and enhance the detection of signals.
Some of the most promising quantum sensor platforms exploit defects in solid-state crystals, such as diamond's negatively charged nitrogen-vacancy (NV) center\,\cite{Barry2020,Shandilya2021}.
Magnetometers based on single NVs enable nanometer-scale spatial resolution\,\cite{Taylor2008, Balasubramanian2008, Maletinsky2012}, while ensembles of NVs have achieved sub-pT/$\sqrt{\text{Hz}}$ direct current (DC) magnetic field sensitivities\,\cite{Barry2024}.
NV magnetometry has enabled applications ranging from biological sensing\,\cite{Steinert2013, Barry2016, Fescenko2019, Kuwahata2020} to magnetic field imaging in condensed-matter systems\,\cite{Maletinsky2012, Fu2014, Glenn2017, Zhou2020, Scheidegger2022, Bopp2025}.
Diamond NVs are particularly suited to sensing at room temperature, where they exhibit long spin-coherence ($T_2\sim2\,\text{ms}$\,\cite{Balasubramanian2008, Maurer2012}) and spin-lattice relaxation times ($T_1\sim6\,\text{ms}$\,\cite{Jarmola2012, Rosskopf2014}). 
Moreover, diamond is mechanically robust and chemically inert\,\cite{Shandilya2022}, enabling sensing in environments inaccessible to trapped-atom platforms\,\cite{Barry2016, Harlander2010} and SQUIDs\,\cite{Fagaly2006}.

\begin{figure}[t!]
    \centering
    \includegraphics[]{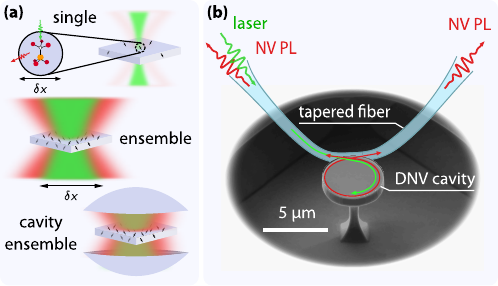}
    \caption{
    \label{fig:1ODMR}\textbf{Magnetometry using NV$^{-}$ centers in a diamond dense-NV cavity.}  
    \textbf{(a)} Different geometries can be used for sensing with varying NV center quantity. A single NV achieves optimal spatial resolution, $\delta x$, whereas ensembles achieve better sensitivity. 
    A cavity-integrated ensemble further enhances sensitivity by making the pump field multi-pass and by improving resonant PL collection (see text for discussion). 
    \textbf{(b)} In this work, a fiber-taper-coupled diamond cavity is used to perform magnetometry. Initialization and readout are achieved using evanescent coupling of light to and from the fiber-taper waveguide. 
    Here, we sketch the fiber-taper atop a scanning electron micrograph of the diamond cavity.
    }
\end{figure}

One key challenge in NV magnetometry is balancing spatial resolution with magnetic sensitivity, as indicated in \Cref{fig:1ODMR}\,(a).
In a magnetometer containing an ensemble of $N$ NVs, the magnetic sensitivity scales as\,\cite{Budker2007}
\begin{align}
    \eta \propto {\frac{1}{ \gamma C \sqrt{N T }} }\,,
    \label{eq:sensGeneral}
\end{align}
where $C$ is the measurement contrast, $T$ is the effective coherence time, and $\gamma/2\pi = 28\,\text{GHz/T}$ is the NV gyromagnetic ratio. 
Advances in readout schemes\,\cite{Shields2015, Jaskula2019, TayefehYounesi2025} and material engineering\,\cite{Balasubramanian2009, Barry2020} have improved both contrast and coherence.
Yet, all systems are limited by a trade-off highlighted by \Cref{eq:sensGeneral}: large ensembles emit more photons, reducing photon-shot-noise and improving sensitivity at the cost of spatial resolution and coherence\,\cite{Bauch2020}; conversely, single defects preserve spatial resolution but are ultimately sensitivity-limited by the photon emission rate and collection efficiency from a single NV\,\cite{Fang2013}. 
An overview of the NV can be found in \Cref{app:NVcenter}.

Tremendous efforts have led to high single-NV collection efficiencies using nanofabricated structures like nanopillars\,\cite{Babinec2010} and solid immersion lenses\,\cite{Hadden2010}, whereas most ensemble magnetometers use free-space optics to collect the NV photoluminescence (PL).
As a result, ensemble magnetometers often have low collection efficiency. 
Although substantial progress has been made in improving ensemble collection efficiency\,\cite{LeSage2012,Wolf2015,Zhang2018,Barry2024} and in fiber-coupling NVs for integrated sensing\,\cite{Strner2021,Graham2023,Li2023,Guo2025}, it remains challenging to simultaneously achieve scalability, low power consumption, high sensitivity, and high spatial resolution.

Here, we address these limitations by integrating high-density NV ensembles into diamond nanophotonic cavities and coupling them to a fiber-taper waveguide used for both NV excitation and collection\,\cite{Masuda2024}.
This compact architecture eliminates bulky optical components, reduces optical power requirements, and provides robust, scalable interfacing via an optical fiber (\Cref{fig:1ODMR}\,(b)).
To our knowledge, this constitutes the first demonstration of a waveguide-coupled nanophotonic cavity-based magnetometer; previous platforms have independently realized cavity~\cite{Katsumi2025} or waveguide-integrated sensors~\cite{Bopp2025,Weng2026}; however, this is the first work to combine the two. 
Using this platform, we demonstrate coherent spin-manipulation and magnetic field detection with a photon-shot-noise-limited DC sensitivity of $58\,\text{nT}/\sqrt{\text{Hz}}$. 

The dense-NV cavity studied here is a diamond microdisk, an example of which is shown in \Cref{fig:1ODMR}\,(b). 
The microdisks were fabricated from a `quantum-grade' single-crystal diamond chip (Element Six, DNV B14)~\cite{ElementSix2021} that maximizes NV concentration ($4.5\,\text{ppm}$) while maintaining long $T_2$, making it well-suited for magnetometry. 
The resulting dense-NV cavities support optical whispering-gallery modes (WGM) with quality factors $Q \sim 10^4$--$10^5$ at visible wavelengths\,\cite{Masuda2024}, and can be evanescently coupled to optical fiber-taper waveguides.
The chip hosts over 200 cavities of varied dimensions, demonstrating scalable fabrication of fiber-addressable devices.

We characterized the optical and spin properties of the dense-NV cavity by measuring PL from cavity NVs using a fiber-taper waveguide (see \Cref{fig:1ODMR}\,(b)).  
Light from a 532\,nm continuous wave (CW) pump laser input to the fiber-taper interacts evanescently with the dense-NV cavity and excites the NVs.
The position of the taper and the polarization of the pump laser were optimized to maximize the fiber-taper-coupled PL intensity.
The fiber-taper enabled efficient excitation, allowing us to perform optimized magnetometry using only 225\,$\upmu$W of pump light.
All measurements were performed on a single cavity with 800\,nm thickness and
4.4~$\upmu$m diameter, selected because it had the highest coupled PL count-rate for a given pump power.
More details of the fabrication, and the electronic and optical setups can be found in \Cref{app:expSetup}.  

\section*{Characterization of the integrated cavity}
\begin{figure}[t]
    \centering
    \includegraphics{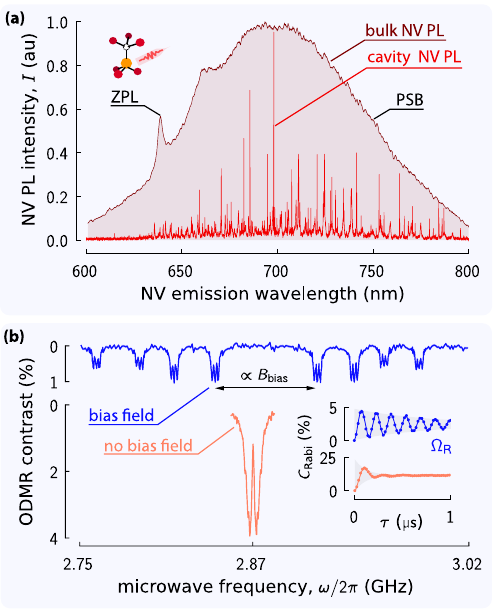}
    \caption{\textbf{Fiber-coupled cavity ODMR.} \textbf{(a)} 
    NV PL collected from a fiber-coupled cavity compared to that collected from the same diamond bulk material using a confocal microscope. 
    The coupled cavity acts as a spectral filter, collecting only photons emitted into resonant modes that couple to the fiber-taper. 
    This is used to enhance the signal-to-noise ratio (SNR) and suppress background PL.
    The most prominent WGM in the range shown has a spectrometer-resolution-limited quality factor of $Q\sim1.3\times10^4$, estimated by a Lorentzian fit to the emission spectrum.
    \textbf{(b)} Optically detected magnetic resonance of NV centers in the cavity as a function of the microwave frequency, $\omega$, with and without a bias magnetic field, $B_{\text{bias}}$. 
    We resolve a hyperfine structure with a linewidth of $1.45~(2)$~MHz and achieve Rabi frequencies up to $\Omega_{\text{R}}/2\pi=6.5$\,MHz (inset). 
    }
    \label{fig:2ODMR}
\end{figure}

\begin{figure*}[b]
    \centering
    \includegraphics[width=\linewidth]{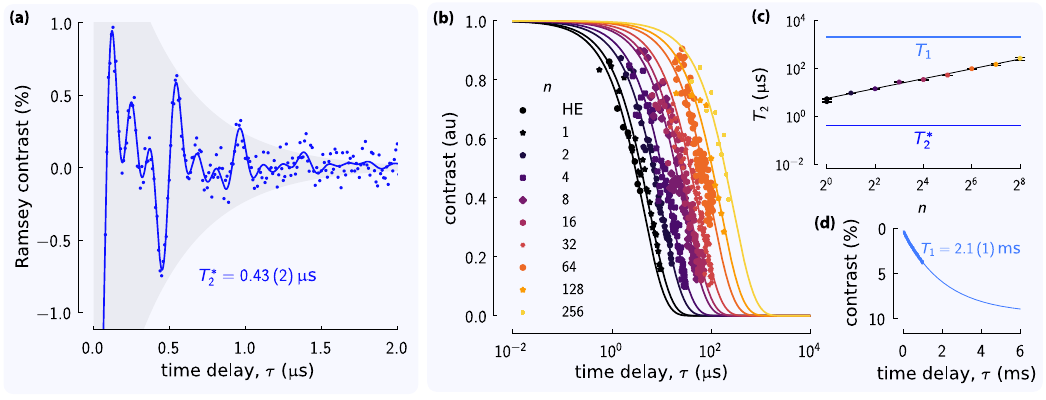}
    \caption{\textbf{Characterizing the spin-coherence time of the diamond microcavity.} \textbf{(a)} The inhomogeneous dephasing time, $T_2^*$, is extracted using Ramsey ODMR. 
    The grey background shows the fitted decay envelope function ($\exp[-\tau/T_2^*]$) of the Ramsey signal (blue points). 
    The blue lineplot is a least squares fit to the data (see \Cref{app:sensitivityOptimization}).
    \textbf{(b)} The spin-coherence time of the cavity is extended using Hahn-echo (HE) and $n$-CPMG pulse sequences, which dynamically decouple the NV centers from magnetic noise. \textbf{(c)} The measured $T_2$ values for each $n$ are compared to $T_2^*$ and $T_1$, which is extracted from the measurement shown in \textbf{(d)}.
    }
    \label{fig:char}
\end{figure*}
Nanophotonic devices tailor the spectrum of light collected from emitters, enhancing or reducing magnetic sensing performance, depending on competing device properties.
Figure \ref{fig:2ODMR}\,(a) compares the normalized NV PL spectrum from the fiber-coupled cavity to that collected from a bulk sample using a confocal microscope.
Only photons emitted into the cavity's WGMs are collected by the fiber-taper, with each peak in the spectrum corresponding to a cavity mode\,\cite{Masuda2024,Flgan2025}.
Numerous WGMs of different families fall within the emission band, giving rise to the many peaks observed in  \Cref{fig:2ODMR}\,(a).
Each of these modes has a unique combination of azimuthal ($m$), radial ($l$) and vertical ($p$) mode numbers. For the fundamental ($l = p = 0$) modes, $m$ is predicted from simulations to vary between approximately $m = 34$ and $m = 44$ across the 637~nm--850~nm NV emission range.
The cavity acts like a spectral filter; photons emitted into the cavity modes are preferentially coupled to the waveguide, while all others are lost.
Approximately $6\,\%$ of the broadband PL emitted by cavity NV centers spectrally overlaps with the collected WGM PL, and we estimate an end-to-end collection efficiency of $\sim 1\,\%$. As discussed in \Cref{app:efficiency}, this end-to-end collection efficiency is estimated from measurements of waveguide-coupling and propagation loss through the system in addition to a comparison of the estimated emission rate of the NV ensemble and the detected optical power.  
Although the present collection efficiency is lower than that achievable using optimized free-space collection optics~\cite{Barry2024}, solid immersion lenses~\cite{Hadden2010,Riedel2014}, or scanning probes~\cite{Hedrich2020}, the present approach provides direct waveguide coupling, eliminating the need for free-space excitation and collection optics.
Importantly, the NV lifetime and subsequent rate of emission are not Purcell-enhanced by the cavity, as room-temperature NV PL is broadband due to its dephasing from phonons, low Debye--Waller factor and a strong phonon-assisted sideband\,\cite{Barclay2011}. 
However, we do expect the cavity to induce some preferential emission into the WGMs\,\cite{Masuda2024,Katsumi2025}.

Despite its limitations, the fiber-coupled cavity provides several advantages.
It enhances signal-to-noise by suppressing stray light collection, enabling measurements in optically noisy environments.
Moreover, only PL from NVs coupled to WGMs contributes, and fiber-coupled green light only excites NVs that spatially overlap with a cavity mode, not those in the bulk diamond.
This ensures that the detected PL originates from a well-defined NV ensemble, whose vertical profile can be smaller than the depth of focus of a conventional microscope objective, as is often done with delta-doped diamond\,\cite{Ohno2012}. 
Therefore, the measured magnetic field can be assigned to a precise location.
In our platform, the vertical and lateral spatial resolution correspond to the thickness ($\sim800\,$nm) and diameter ($\sim4.4\,\upmu$m) of the disk, respectively.

The benefits of using fiber-coupled nanophotonic cavities also extend to microwave control and the magnetic sensitivity.
The DC sensitivity of an NV magnetometer scales with the inhomogeneous dephasing time, $T_2^*$, which degrades under spatial inhomogeneities in the microwave field and the bias magnetic field~\cite{Barry2020}. 
Because the cavity defines a small sensing volume (12\,$\upmu$m$^3$),
it is easier to generate spatially homogeneous microwave pulses across the NV ensemble, thereby supporting longer $T_2^*$.
Further details are provided in \Cref{app:expSetup}. 

Using the apparatus, we measure optically detected magnetic resonance (ODMR) spectra using a pulsed scheme, both with and without an external bias magnetic field (\Cref{fig:2ODMR}\,(b)).
Without a bias field, we achieve 18\,\% NV PL contrast by resonantly driving all electronic spins with a microwave pulse with Rabi frequency $\Omega_{\text{R}}/2\pi = 6.5\,\text{MHz}$.
This contrast is lower than the empirical
maximum of 30~\%\,\cite{Manson2006}, which we attribute to difficulties in simultaneously polarizing all four NV crystallographic orientations, temporal imperfections in the microwave pulses, and fiber PL generated by the green laser interacting with dopants in the fiber core. Note that the effect of fiber PL is detectable but small; the detected NV PL is over two orders of magnitude more intense than the detected fiber PL (\Cref{app:opticalSaturation}).
Therefore, the mitigation of background noise we achieve through fiber coupling is not spoiled by the presence of fiber PL.
Furthermore, at lower microwave power the bias-free spectrum reveals splitting caused by local electric fields from charged defects, enabling us to verify the density of NV ($4.5\,\text{ppm}$) and substitutional nitrogen ($8.5\,\text{ppm}$) defect centers.
The absence of splitting or broadening of the zero-field ODMR spectrum beyond those caused by the nominal defect concentrations suggests that fabrication-induced strain has negligible effect on the NV spin states\,\cite{Ruf2019,Guo2021} (\Cref{app:defectConcentration}). 
With a bias magnetic field $\mathbf{B}_{\text{bias}}$ applied, each NV orientation can be addressed individually: the eight observed resonances correspond to the $|m_s = 0\rangle \rightarrow |m_s = \pm 1\rangle$ transitions of the four crystallographic orientations. 
The bias field also suppresses broadening from electric-field and strain inhomogeneities\,\cite{Kehayias2019}, yielding a well-resolved hyperfine structure.
From these measurements, we extract a hyperfine linewidth of $\nu = 1.45\,(2)\,\text{MHz}$.

To better assess the extent of fabrication-induced strain, we measure $T_2^*$ using Ramsey ODMR (\Cref{fig:char}\,(a)), obtaining $T_2^* = 0.43\,(2)\,\upmu$s.
This estimate is slightly shorter than the reported $T_2^*$ for the bulk diamond we use to make the cavities (500\,ns)\,\cite{ElementSix2021}. 
Though some nanofabrication processes can induce strain and degrade spin-coherence\,\cite{Ruf2019,Guo2021}, the small reduction we observe---significantly less than that observed in previous nanofabricated diamond cavity sensors\,\cite{Katsumi2025}---supports the conclusion that our fabrication process introduces only a small amount of strain to NVs within the sensing volume.
The relatively short $T_2^*$ is primarily limited by dephasing due to surrounding paramagnetic impurities, such as N$_{\text{s}}^0$ defects.
As demonstrated by prior work, $T_2^*$ can be extended by more than an order of magnitude through spin-bath driving\,\cite{DeLange2012,Knowles2013,Bauch2018}, an approach that is especially useful when combined with double-quantum magnetometry\,\cite{Fang2013}.
In our case, however, the baseline $T_2^*$ is too short for double-quantum magnetometry to yield a clear advantage on its own\,\cite{Barry2020}.

To achieve longer coherence times, we employ dynamical decoupling sequences to mitigate both dephasing and decoherence processes.  
As shown in \Cref{fig:char}\,(b) and (c), we use a Carr–Purcell–Meiboom–Gill (CPMG) protocol\,\cite{Meiboom1958} with varying refocusing-pulse quantities, $n$.
Using $n=256$, we extend the coherence time by a factor of $\sim65$ compared to a Hahn-echo (HE) baseline ($T_2^{\text{HE}} = 4.2\,(3)\,\upmu$s\,$\rightarrow\,T_2^{\text{(256)}} = 0.27\,(4)\,$ms), approaching the longitudinal spin-relaxation limit, $T_1 = 2.1\,(1)\,\text{ms}$ (see \Cref{fig:char}\,(d)).
Further details on the pulse sequences are provided in \Cref{app:measSequences}.

\section*{Magnetometry demonstration}

\begin{figure*}[ht!]
        \centering
        \includegraphics[width=\linewidth]{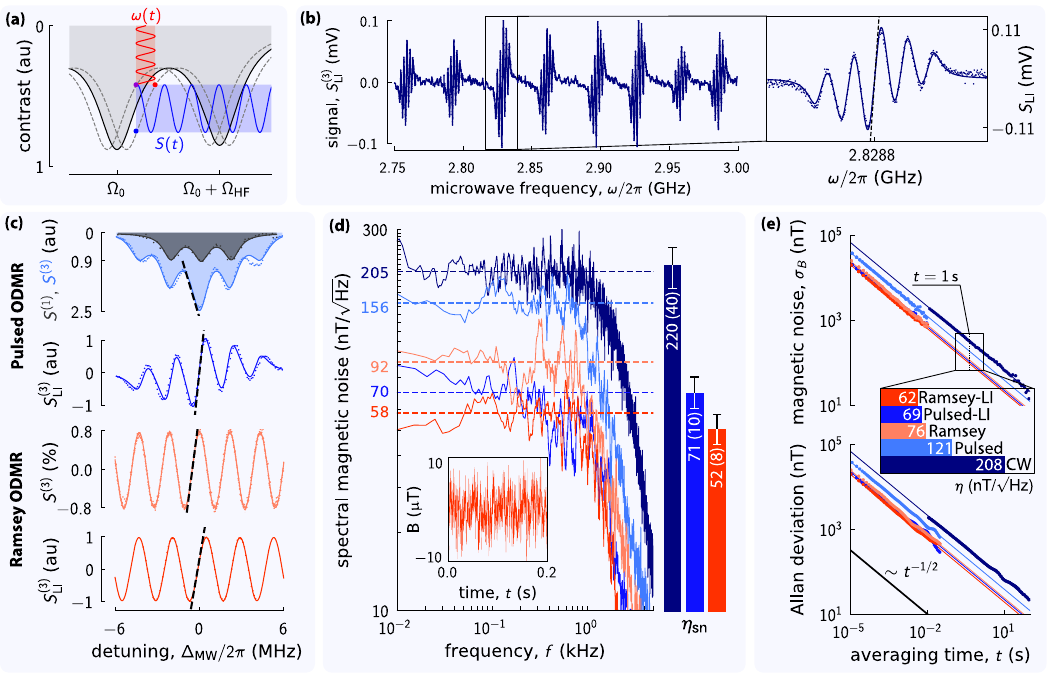}
        \caption{
        \textbf{Magnetometry using the microcavity sensor.}
        \textbf{(a)} Lock-in techniques suppress electronic and laser noise. 
        Frequency-modulated microwaves (red) cause the NV PL signal (grey) to fluctuate at the modulation frequency, which can be extracted to suppress noise.
        \textbf{(b)} Measurement of NV ensemble spin transitions split by a bias magnetic field using lock-in detection. The dashed black line indicates the operating point for maximum signal response.
        \textbf{(c)} ODMR spectra comparing single-tone (grey) and three-tone (blue) microwave driving. Simultaneous driving of all three $^{14}$N nuclear hyperfine transitions ($\omega, \omega \pm \Omega_{\text{HF}}$) increases signal contrast and results in five spectral features; $\Omega_{\text{HF}}$ is the $^{14}\text{N}$ hyperfine splitting ($|A_{\parallel}|$, \Cref{table:modelParams}). 
        Lock-in measurements done for both the pulsed and Ramsey ODMR protocols produce similar contrast to their non-lock-in counterparts; however, they suppress low-frequency noise and result in higher sensitivity. 
        \textbf{(d, e)} Magnetic field measurements and Allan deviation for different sensing schemes. The inset in \textbf{(d)} shows 200\,ms of Ramsey-LI data. 
        Lock-in pulsed ODMR and Ramsey measurements are shot-noise-limited, as demonstrated by the bar chart in \textbf{(d)} which shows the predicted shot-noise-limited sensitivities ($\eta_{\text{sn}}$).
        The sensitivities measured in the data presented in \textbf{(d)} and \textbf{(e)} are in good agreement. 
        The similarities between the Allan deviation and the magnetic noise signify the sensor's stability---it is not limited by drift. 
        }
        \label{fig:sens}
    \end{figure*}
    
We next demonstrate magnetic field sensing with the dense-NV cavity using both CW and pulsed protocols, each tailored for different experimental needs, and show that noise-suppression strategies enable photon-shot-noise-limited sensitivity.
These results establish the dense-NV cavity magnetometer as a flexible platform suited to a wide range of sensing applications.

CW-ODMR is commonly employed for NV magnetometry because of its simplicity.
In this approach, a CW laser pump and CW microwave drive are applied simultaneously while sweeping the microwave frequency to identify NV spin transitions. 
This method is limited by power broadening and electronic noise; the latter can be mitigated using lock-in (LI) detection: by modulating the microwave frequency at frequency $\omega_{\text{LO}}$ and extracting the signal at that frequency, noise outside the modulation bandwidth is suppressed and magnetometer sensitivity is improved (see \Cref{fig:sens}\,(a)).
An example CW-ODMR spectrum obtained from our device is shown in \Cref{fig:sens}\,(b), measured using LI detection to suppress noise and improve sensitivity.

To reach the sensitivity limit of the cavity, it is necessary to implement pulsed ODMR and Ramsey protocols.
Protocols that pulse the optical and microwave fields can achieve higher sensitivity than CW magnetometry because stronger fields can be applied without introducing significant power broadening.
All pulsed measurements in this work employ three-tone microwave pulses that simultaneously address the three $^{14}$N hyperfine transitions, enhancing signal contrast by a factor of $\sim2.8$.
This effect is illustrated in \Cref{fig:sens}\,(c), where the single-tone case is compared with three-tone driving---the three-tone signal, $S^{(3)}$, produces five spectral features compared to the three produced by a single-tone signal, $S^{(1)}$. 
To improve Ramsey readout fidelity, we also adapt the lock-in detection scheme used for pulsed ODMR\,\cite{Zhang2022}. 
To the best of our knowledge, this is the first time such a lock-in technique has been used to improve the readout fidelity of an NV Ramsey scheme.
Details of all pulse sequences and optimization procedures are provided in \Cref{app:measSequences,app:sensitivityOptimization}, respectively.

To determine the optimum operating point of each protocol, we record the PL strength for varying microwave detuning $\Delta_{\text{MW}} = \omega - \Omega_0$, relative to the spin transition at frequency $\Omega_0$.
The resulting spectra for pulsed and Ramsey ODMR, along with their lock-in counterparts, are shown in \Cref{fig:sens}\,(c). 
For each scheme, we identify the frequency where the signal response is maximally sensitive to changes in the magnetic field, as indicated by dashed lines in \Cref{fig:sens}\,(c).

With the system operating at the point of maximum sensitivity, magnetic field sensitivity is then determined using two complementary methods.
In the first approach, a signal $S$, recorded as a function of time, is converted to magnetic field, $B=S(\gamma\partial S/\partial \Delta_{\text{MW}})^{-1}$. The corresponding frequency spectrum of the time series, shown in \Cref{fig:sens}\,(d) for both CW-ODMR and the pulsed schemes, provides sensitivities given by the average noise within the effective noise bandwidth of the spectrum (dashed lines). 
In the second approach, magnetic field signals from each protocol are recorded and the sampling time $t$ used to measure the magnetic field is varied. 
The resulting standard deviation $\sigma_B$ of each time series is plotted in \Cref{fig:sens}\,(e), whose fit provides an independent estimate of the sensitivity in agreement with the previous method. 
We measure the magnetic noise to be less than 20\,nT for the CW protocol, given a sampling time of 100 seconds. 

The lower panel of \Cref{fig:sens}\,(e) shows the Allan deviation, which characterizes the stability of the magnetometer and demonstrates that the magnetometer can be used to perform high-sensitivity measurements given longer acquisition times. 
Similarities between the Allan deviation and the magnetic noise demonstrate that the sensor is noise- and not drift-limited.
Stable operation over long timescales is crucial for measurements requiring signal integration or repeated scans of nanoscale fields. 
Finally, note that in \Cref{app:testField} we directly apply a test magnetic field to demonstrate the cavity's usefulness as a sensor.

The most sensitive protocol is the lock-in Ramsey method, with a sensitivity of 58\,nT/$\sqrt{\text{Hz}}$.
Across all methods, the sensitivities follow the expected trend: lock-in pulsed measurements outperform non-lock-in pulsed measurements, which in turn surpass the CW-based approach. 
Crucially, all lock-in measurements are photon-shot-noise-limited, as demonstrated by the bar chart in \Cref{fig:sens}\,(d); the measured sensitivities of these three protocols are within uncertainty of the predicted shot-noise-limited sensitivities, $\eta_{\text{sn}}$, indicating that the magnetometer’s sensitivity is fundamentally constrained by photon statistics rather than electronics. 
Calculations on the photon-shot-noise limits can be found in \Cref{app:sensitivityOptimization}.

\section*{Sensor performance and applications}

    \begin{figure}[t]
    \centering
    \includegraphics[]{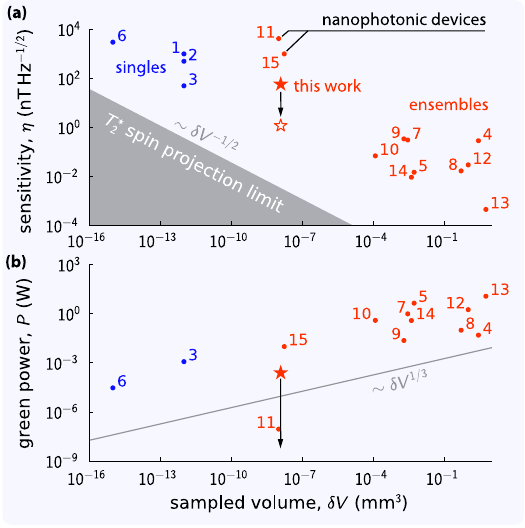}
    \caption{
    \textbf{Comparison of the figures of merit of different diamond magnetometers.} 
    Comparison of the sensitivity \textbf{(a)} and optical control power \textbf{(b)} as a function of volumetric spatial resolution for different magnetometers composed  of single and ensembles of NV centers: 1\,\cite{Taylor2008}; 2\,\cite{Maze2008}; 3\,\cite{Fang2013}; 4\,\cite{Clevenson2015}; 5\,\cite{Barry2016}; 6\,\cite{Pelliccione2016}; 7\,\cite{Patel2020}; 8\,\cite{Zhang2021}; 9\,\cite{Strner2021}; 10\,\cite{Shim2022}; 11\,\cite{Li2023}; 12\,\cite{Graham2023}; 13\,\cite{Barry2024}; 14\,\cite{Sekiguchi2024}; 15\,\cite{Katsumi2025}. 
    For single NV sensors, $\delta V$ is approximated by either the experimental spatial resolution or the resolution of the confocal microscope.
    The black arrows and outlined star highlight the sensitivity improvement estimated given the modifications outlined in the main text.
    The $T_2^*$-limited spin projection limit is calculated as $\eta_{\text{sp}}= (2\pi\gamma \sqrt{NT_2^*})^{-1}$, where $N$ is the sample volume multiplied by the NV density, $\text{[NV]}=4.5\,$ppm.
    }
    \label{fig:litReview}
    \end{figure}
    The spatial resolution and magnetic sensitivity of NV-based sensors are intrinsically linked, as discussed above, and the optimal device architecture depends strongly on the target application.
    Figure \ref{fig:litReview}\,(a) compares the performance of the dense-NV cavity sensor reported here with several other single-NV and ensemble-NV magnetometers\,\cite{Taylor2008,Maze2008,Fang2013,Clevenson2015,Barry2016,Pelliccione2016,Patel2020,Strner2021,Zhang2021,Shim2022,Li2023,Graham2023,Barry2024,Sekiguchi2024,Katsumi2025}, as well as the spin-projection-limit of the diamond used here. 
    Although the present device neither exceeds the ultimate sensitivity of large ensembles, nor the spatial resolution of a single NV, it occupies an otherwise sparse region in the performance landscape and achieves the highest sensitivity reported to date for an on-chip nanophotonic cavity magnetometer\,\cite{Li2023,Katsumi2025}.
    Improvements over previous works are attributed to several advantages of our platform, including the high density of NV centers and the waveguide collection of PL.
    Even though only a fraction of the NV PL is emitted into WGMs, those photons can be efficiently routed into the waveguide, enabling integrated excitation and readout without free-space optics (\Cref{app:efficiency}).
    Additionally, the device requires low optical pump power (see \Cref{fig:litReview}\,(b) and \Cref{app:expSetup}), which benefits scalability and efficiency\,\cite{Yu2024,Solyom2024}---an advantage in sensitive environments where intense optical fields can damage samples\,\cite{Blzquez-Castro2019}.  
    The non-resonant device described in \cite{Li2023} achieved low power operation by capitalizing on the high fiber-coupling efficiency of the pump field to their device. 
    Therefore, despite the already low pump power, the power consumption of the cavity sensor studied here could be made even lower by improving the coupling between the waveguide and the cavity at the pump wavelength and by engineering a better cavity mode at the pump wavelength (see \Cref{app:expSetup}).

    Furthermore, the sensor is fabricated on-chip and the NV PL excitation and collection is evanescently fiber-coupled, enabling straightforward integration with photonic components and on-chip microwave delivery.
    This architecture is compatible with waveguide-coupled photonic circuits and scalable to two-dimensional sensor arrays, creating new opportunities for applications in biological sensing\,\cite{Krber2016,Barry2016} and current mapping\,\cite{Zhou2020}. 
    While certain applications require sub-nT$/\sqrt{\text{Hz}}$ sensitivities, the present sensor does not reach this level. 
    Nonetheless, targeted modifications make sub-nT$/\sqrt{\text{Hz}}$ sensitivities within reach.
        
    The dense-NV cavity's sensitivity is limited by several factors: (i) low total collection efficiency ($\sim1\,\%$ - see \Cref{app:efficiency}); (ii) fiber-PL-limited readout fidelity; and (iii) inhomogeneous dephasing (low $T_2^*$) caused by other electronic and nuclear spins in the diamond~\cite{Bauch2020}. 
    Engineering the WGM cavity to have a broader linewidth and higher mode density, and coupling it to a hybrid photonic waveguide\,\cite{Neshasteh2025} in aluminum nitride\,\cite{Wan2020} or silicon nitride\,\cite{Katsumi2025} would significantly improve the collection efficiency and mitigate fiber PL. 
    Further, although the cavity exhibits features of some resonant enhancement of the green laser, its geometry was not optimized for that purpose; designing for optimal pump resonance would further boost the efficiency and sensitivity of the magnetometer. 
    
    Sensitivity can also be improved by extending $T_2^*$ through control of the spin environment. 
    Spin-bath-driving protocols\,\cite{DeLange2012,Knowles2013,Bauch2018}, higher conversion efficiency of substitutional nitrogen to NV \cite{Balasubramanian2009}, and reduced $^{13}$C concentration \cite{Bar-Gill2013}, could together yield another order-of-magnitude improvement.
    Further increases in NV density could also reduce photon-shot-noise at the cost of shorter $T_2^*$.
    Conversely, reductions in NV density would result in longer $T_2^*$; however, the subsequent increase in photon-shot-noise would be more detrimental to the sensitivity than the improvements in the ODMR linewidth\,\cite{Barry2020}.
    With these combined enhancements, the sensitivity could be improved by nearly two orders of magnitude and could realistically achieve sub-nT DC sensitivities, as indicated by the projection in \Cref{fig:litReview}\,(a) allowing our platform to achieve state-of-the-art readout fidelity\,\cite{Barry2024}.
    The platform is also compatible with AC magnetometry, which could benefit from the extended coherence times reported here.

    While the present work demonstrates a proof-of-concept fiber-coupled platform, integration with on-chip waveguides\,\cite{Wan2020,Neshasteh2025} could enable scalable sensors for a variety of future applications.
    For one, diamond biomagnetic sensors are of great interest\,\cite{Krber2016,Barry2016} and have been used to measure electric currents in the heart\,\cite{Arai2022} and nervous tissue\,\cite{Barry2016}, often requiring micrometer-scale spatial resolution\,\cite{Sekiguchi2024}.
    While existing NV sensors are limited in either spatial resolution or integration potential, the present platform could enable proximal measurements of microscopic biological samples that are otherwise inaccessible.
    Samples need not be permanently deposited on the sensor surface. Instead, targets could be positioned near the cavity using nano-positioners~\cite{Rugar2014,Hberle2015} or microfluidic channels~\cite{Steinert2013}.
    The microdisks are mechanically robust, routinely surviving repeated wet-processing steps, flowing liquids, and nitrogen blow drying during fabrication and handling. This robustness suggests compatibility with future microfluidic applications.
    The dense-NV cavities might also be useful for nuclear magnetic resonance (NMR) spectroscopy. 
    Bulk diamond sensors have probed the NMR spectra of molecular\,\cite{Glenn2018,Liu2022,Briegel2025} and biological\,\cite{Neuling2023} samples, enabling detection of picoliter volumes\,\cite{Glenn2018} under modest bias magnetic fields $(<1\,\text{T})$.
    The high NV density of the cavities could enhance NMR sensitivity in compact geometries suitable for specialized biological or medical analyses, while advanced AC sensing protocols can provide spectral resolution beyond the $T_2$-limit\,\cite{Boss2017, Schmitt2017}.
    Importantly, nanofabrication and the high defect density do not substantially compromise the measured spin coherence times, preserving the long $T_1$ and $T_2$ values required for sensitive AC magnetometry and spectroscopic applications.
    Finally, NV-based magnetometers are increasingly used for two-dimensional magnetic imaging, whether through scanning\,\cite{Maletinsky2012,Thiel2019,Zhou2020,Welter2022,Scheidegger2022,Li2023} or pixelated detection\,\cite{Bopp2025}. 
    Existing platforms, however, are often limited in speed, spatial resolution, or scalability.
    Arrays of dense-NV cavities~\cite{Barclay2006} integrated with photonic waveguides~\cite{Wan2020} and lithographically defined microwave antennas~\cite{Knaut2024} could allow imaging of microscopic magnetic field gradients with high sensitivity. 
    Integration with pixelated flux concentrators\,\cite{Fescenko2020} could further improve performance and lead to applications from current mapping to magnetic domain measurement\,\cite{Hedrich2021}.  
    
    Realization of these applications would require advances in device packaging, including replacement of the fiber-taper with permanently integrated on-chip waveguides. While the fiber-taper provides a flexible interface for proof-of-concept experiments, practical deployment and proximal sensing of nearby samples will likely demand on-chip waveguides.
    
\section*{Conclusions}

We have demonstrated an on-chip fiber-coupled magnetometer based on a whispering-gallery-mode dense-NV diamond cavity, whose dense ensemble of NVs is evanescently excited and measured with a fiber-taper waveguide to perform magnetometry. 
The cavity’s microscopic dimensions enable high sensitivity in a small footprint.
With lock-in-amplified Ramsey ODMR measurements, we achieve a photon-shot-noise-limited static field sensitivity of 58\,nT/$\sqrt{\text{Hz}}$, benchmarked against several magnetometry schemes.
Using dynamical decoupling, we measure the coherence time of the sensor, whose magnitude ($T_2 = 0.27\,(4)\,$ms) motivates the platform's use in AC sensing applications.
This work represents the first waveguide-coupled, on-chip NV cavity magnetometer.
It is laser-power-efficient, it achieves the highest sensitivity among nanofabricated cavities, and it advances the state-of-the-art of nanofabricated NV magnetometers.
The combination of compact size, integration potential, and robust photon-shot-noise-limited performance positions these dense-NV cavities as versatile tools for next-generation quantum-sensing applications. 

\subsection*{Data availability}
The data supporting the findings of this study are found within the main text and the appendices. 
Source data and code may be obtained from the authors upon reasonable request.

\subsection*{Acknowledgements}
We gratefully acknowledge the time and expertise of the staff at the Nanofab at the University of Alberta during device fabrication. 
The equipment lent to us by Peter Gimby and the Electronics Workshop team at the University of Calgary was greatly helpful. 
Lastly, we sincerely thank Joe Losby and Joe Itoi for fruitful discussions.
This work was supported by NSERC (Discovery Grant program), Alberta Innovates, and the Canada Foundation for Innovation. SF acknowledges support from the Swiss National Science Foundation (Project No. P500PT\_206919).

\subsection*{Author contributions}
N.J.S. performed all spin-characterization and magnetometry measurements with assistance from V.K.K. and S.F. 
E.Z. fabricated the cavities, and V.K.K. characterized the bulk spectral emission. J.S.W. simulated and optimized the design of the microwave antenna.
N.J.S. performed all data analysis and simulation and prepared the manuscript with feedback from all authors. 
P.E.B. guided and supervised the work. 

\clearpage

\newpage

\appendix

\section*{Appendix}
\section{The diamond NV$^{-}$ center}
\label{app:NVcenter}

    \begin{figure}[b!]
    \centering
    \includegraphics[]{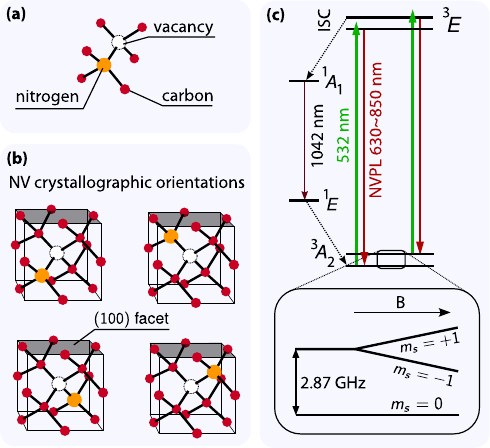}
    \caption{
    \textbf{The NV center in diamond.} 
    A schematic shows the arrangement of the NV center within the diamond lattice \textbf{(a)}, as well as the four NV crystallographic orientations in the $(100)$-oriented diamond sample \textbf{(b)}. \textbf{(c)} The energy level diagram of the negatively charged NV center, which highlights the ground state Zeeman splitting of the electronic spins.  
    }
    \label{fig:SI_NV_center}
    \end{figure}
The negatively charged nitrogen–vacancy (NV) center in diamond consists of a substitutional nitrogen atom adjacent to a lattice vacancy, with an additional electron supplied by a nearby donor (see \Cref{fig:SI_NV_center})\,\cite{Doherty2013}.  
Its ground state is a spin triplet ($^3A_2$), with sub-levels $|m_s=0\rangle$ and $|m_s=\pm1\rangle$ separated by the zero-field splitting $D \approx 2.87$~GHz\,\cite{Manson2006}.  
An applied magnetic field can lift the degeneracy of the $|m_s=\pm1\rangle$ sub-levels via the Zeeman effect, enabling magnetic-field-dependent spin transitions.  
Optical excitation promotes the population to a spin triplet excited state ($^3E$), which decays radiatively and produces stable visible PL with a zero-phonon line at 637\,nm (see \Cref{fig:2ODMR}\,(a)).  
Crucially, optical excitation of the NV center is spin-conserving, even for far off-resonant excitation.
In addition to the radiative decay channel, there exists a spin-selective intersystem crossing (ISC) pathway from the excited triplet to a metastable singlet state ($^1A_1$).
The singlet state ($^1E$) is long lived, and it `shelves' NV centers from optical cycling when they transition through it.
The ISC is more likely from the $|m_s=\pm1\rangle$ states, which shelves them and results in reduced PL intensity when these states are occupied. 
Further, repeated cycling through the ISC eventually results in polarization to the $|m_s = 0\rangle$ state.
This mechanism produces both optical spin polarization and spin-dependent PL contrast, which ODMR exploits to measure spin transitions\,\cite{Barry2020}.  
After initialization into $|m_s=0\rangle$, a resonant microwave field drives the population into $|m_s=\pm1\rangle$.  
Upon spin-conserving optical excitation, the population in the excited $|m_s=\pm1\rangle$ state is more likely to decay through the ISC resulting in reduced PL. 
Therefore, sweeping the microwave frequency across a spin resonance produces a dip in PL at resonance, forming the basis of ODMR.  

\section{Experimental methods}
\label{app:expSetup}

\begin{figure*}[t!]
    \centering
    \includegraphics[width=\linewidth]{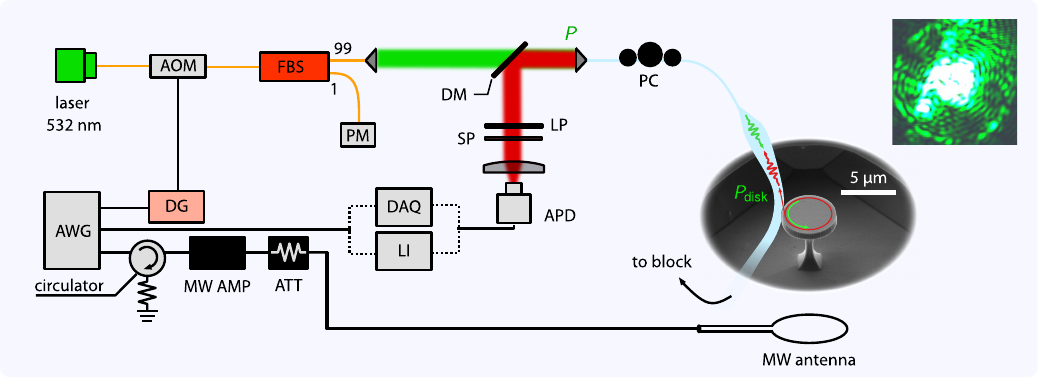}
    \caption{\textbf{Experimental setup.} 
    A green laser is coupled into free-space and filtered before being fiber-coupled and evanescently coupled to the dense-NV cavity sensor. 
    The NV PL is coupled into the fiber-taper and the counter-propagating PL is filtered in free-space before being detected. 
    Microwave electronics are used to control the spin state of the NV ensemble and the measurement timing.
    An optical microscope image shows the green laser scattering from the cavity (top right). 
    The scattering is strongly dependent on polarization, suggesting that the green laser couples to a cavity mode.
    $P$ is the green power input to the fiber-taper, and $P_{\text{disk}}$ is the power that reaches the tapered region.
    List of abbreviations: AOM --- acousto-optic modulator; AMP --- amplifier; APD --- avalanche photodiode; ATT --- attenuator; AWG --- arbitrary wave generator; DAQ --- data acquisition tool; DG --- delay generator; DM --- dichroic mirror; FBS --- fiber beamsplitter; LI --- lock-in amplifier; LP --- long pass filter; MW --- microwave; PC --- polarization controller; PM --- power meter; SP --- short pass filter.
    }
    \label{fig:SI_Setup}
    \end{figure*}

The experimental setup used to characterize the dense-NV cavity sensors is shown schematically in \Cref{fig:SI_Setup}.
\label{sec:SI-setup}
\subsection{Fabrication of the microdisks}
The diamond dense-NV cavities were etched from a $( 100 )$-oriented quantum-grade single-crystal diamond substrate grown using chemical vapor deposition (Element Six, DNV B14)~\cite{ElementSix2021}.
The diamond sample contains a dense ensemble of NVs with a nominal concentration of 4.5~ppm and pre-treatment substitutional nitrogen concentration of 13~ppm.
Prior to fabrication, the sample was mechanically polished, then a previously reported quasi-isotropic undercut method was used to create the cavities~\cite{Khanaliloo2015NanoLett, Mitchell2019}.
Compared to optical-grade diamond\,\cite{ElementSix2021CVD}, the quantum-grade diamond required $30$--$40\,\%$ longer oxygen etch times, which we attribute to the different concentration of crystal defects~\cite{Khokhryakov2015,Inyushkin2023}. 
Although we did not perform pre-fabrication etching of the diamond surface, it could have been used to reduce subsurface damage and strain caused by mechanical polishing~\cite{Hicks2019}, which could have extended the inhomogeneous dephasing time of the NV ensemble.
After fabrication, the cavities were cleaned in a piranha solution (3:1 H$_2$SO$_4$:H$_2$O$_2$), followed by a boiling tri-acid (equal parts concentrated sulfuric, nitric, and perchloric acids) solution to remove surface graphite and remnant nanofabrication residues\,\cite{Brown2019}. The tri-acid cleaning procedure primarily oxygen-terminates the sample.
The final fabricated chip contained over 200 dense-NV WGM cavities of various dimensions, engineered to host high quality-factor optical WGMs.

\subsection{Optical setup}
Control and readout of the NV centers is performed using a CW 532\,nm laser (CrystaLaser LC CL532-025-SO), which is evanescently coupled to the cavity using a dimpled fiber-taper. 
First, the laser is fiber-coupled and injected into an acousto-optic modulator (AOM, AA Opto-electronic MT200-BG18) for amplitude modulation.
The modulated light is then sent to a 99:1 fiber beam splitter (Thorlabs TW560R1A1), which directs 1~\% of the light to a power meter (Thorlabs PM400/S120C). 
The other 99~\% is coupled into free-space, where it passes through a dichroic mirror (Thorlabs DMLP567) and is fiber-coupled into a dimple-tapered optical fiber~\cite{Michael2007,Masuda2024} (SMF-28). 
The 532\,nm light evanescently scatters into the coupled dense-NV cavity via the fiber-taper, and the generated NV PL is coupled from the cavity into the fiber-taper in both directions, both co- and counter-propagating with the pump.
The counter-propagating NV PL is coupled into free-space, collimated, then filtered from the green pump light using the dichroic mirror, a 633\,nm long pass filter (BLP01-633R-25), and a 940\,nm short pass filter (FF01-940/SP-25), before being focused onto an avalanche photodiode (Thorlabs APD410A/M). 

The reason we collect the counter-propagating PL rather than the forward-propagating PL is the presence of collinear Raman processes in the fiber, which travel in the same direction as the pump laser.
For further discussion of fiber PL and defect saturation, see \Cref{app:opticalSaturation}.

\subsection{Cavity fiber-coupling}
\label{app:cavityFiberCoupling}
The chip was placed on a microwave antenna printed circuit board (PCB), mounted atop a two-axis ($x$-$y$) stepper-motor translation stage (Suruga Seiki XXC06020-G). 
For details on the design and characterization of the PCB, see \Cref{sec:SI-microwave}.
Green light and NV PL were evanescently coupled between the fiber-taper and the dense-NV cavity by attaching the fiber-taper to a third stage ($z$) and positioning the dimpled region of the taper near to the cavity. 
The taper was then brought into contact with the side of the disk, choosing the position to coincide with a maximum NV PL signal.
Once in contact, the taper remained stable, ensuring constant fiber–cavity coupling throughout the experiment.
The nature of the cavity also helps to overcome mode mismatch between the diamond and the fiber-taper waveguide.
To avoid undesired contact between the fiber-taper and the diamond substrate, the fiber-taper is fabricated with a dimple\,\cite{Michael2007,Masuda2024}.

    \begin{figure*}[hb!]
    \centering
    \includegraphics[width = \textwidth]{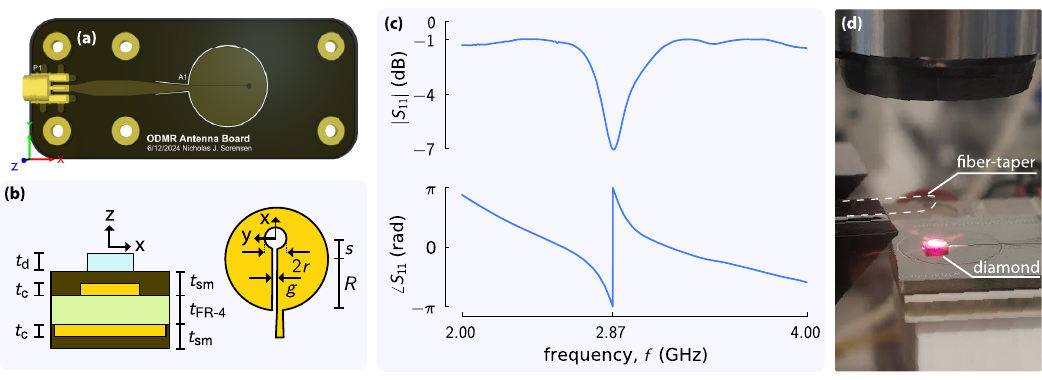}
    \caption{\textbf{The planar ring microwave antenna used to perform ODMR. }
    \textbf{(a)} Rendered image of the printed circuit board.
    \textbf{(b)} Dimensions of the board with minor modifications, adapted from Sasaki et al.\,\cite{Sasaki2016}: $R=7.0\,\text{mm}$, $r=0.5\,\text{mm}$, $s=3.9\,\text{mm}$, $g=0.1\,\text{mm}$.
    The board area measures $60\times 25\,\text{mm}^2$ and is 1.6\,mm thick.
    Each copper layer is $t_{\text{c}}=0.035\,\text{mm}$ thick and the FR4 dielectric is $t_{\text{FR4}}=1.53\,\text{mm}$ thick. 
    The diamond is $t_{\text{d}}=0.5\,\text{mm}$ thick, and the solder mask is $t_{\text{sm}}=0.03\,\text{mm}$ thick. 
    \textbf{(c)} The measured $S_{11}$ parameter aligns well with that predicted by a simulated model---the resonance frequency sits at 2.87\,GHz with a return loss of 7\,dB on resonance. 
    \textbf{(d)} Photograph of the diamond chip sitting atop the antenna PCB. 
    The highlighted fiber-taper hovers above the chip underneath the center of the microscope objective. 
    }
    \label{fig:SI_MWAntenna}
    \end{figure*}
Although we could not directly measure transmission spectra of the cavity’s visible modes due to the absence of a tunable source across the NV$^{-}$ PL spectrum and the difficulty of performing transmission spectroscopy at visible wavelengths in the telecom fiber-taper, several observations indicate coupling of the green laser into a cavity mode:  
(i) the high density of optical modes near 532\,nm (an assumption, given the high density at NV PL wavelengths seen in \Cref{fig:2ODMR}\,(a) in conjunction with the trend that mode density increases with frequency); 
(ii) strong green scattering observed from the cavity (see \Cref{fig:SI_Setup}), consistent with large circulating power;  
(iii) both the scattered light and NV PL were sensitive to temperature and pump polarization. 
These features are consistent with resonant coupling of the pump to a cavity mode, though none provide conclusive evidence or allow measurement of the quality factor and corresponding field enhancement of the mode.
Despite the observation of scattering into the device, the coupling of green light into the fiber is very weak and most of the green light just transmits through the fiber-taper.
Therefore, great improvements in power usage can be achieved by engineering the cavity to have a high-$Q$ mode at the pump wavelength and by using a fiber-taper with a narrower tapered region\,\cite{Masuda2024}.

To maximize the coupling of green light from the fiber into the disk, the temperature of the disk was tuned between 20$\degree$C and 30$\degree$C using a thermoelectric cooler, across which the detected PL varied by approximately 50\,\% from the maximum. 
The detected NV PL signal is maximized at 24$\degree$C, at which temperature we performed all measurements.
The amount of green light scattered into the dense-NV cavity was also polarization dependent; we maximized the scattering using a fiber polarization controller.

\subsection{Microwave electronics and antenna design}
\label{sec:SI-microwave}
An arbitrary waveform generator (AWG, Tektronix AWG70002A) was used to control the AOM via a delay generator (Stanford DG535), trigger the data acquisition tool (Tektronix DSA70804B) or lock-in amplifier (Zurich HF2LI), depending on the application, and produce the microwave pulse sequences. 
The microwave pulses sent from the AWG were amplified (Mini-Circuits ZHL-25W-63+) and attenuated before being sent to the microwave PCB antenna. 
The AWG was protected from any back-reflected power from any impedance mismatching using a circulator.

The antenna used to deliver the microwave pulses, as shown in \Cref{fig:SI_MWAntenna}\,(a), is a modified version of the one described by Sasaki et al.\,\cite{Sasaki2016}. 
It comprises a microwave planar ring antenna with a resonance frequency of 2.87\,GHz and a bandwidth of around 400\,MHz. The geometric parameters of the board are defined in \Cref{fig:SI_MWAntenna}\,(b).
After assembly, the board was modified to fit atop the stepper-motor translation stage and to adjust the antenna resonance frequency.
One side of the antenna PCB was trimmed off to align the center ring of the antenna with the fiber-taper.
Post modification, we characterized the $S_{11}$ parameter and found the resonance frequency to be slightly too low $(\sim2.8\,\text{GHz})$.
The resonance frequency of the antenna can be approximated by the proportionality~\cite{Sasaki2016}
\begin{align}
    f_0 \propto   \sqrt{\frac{g}{r(R+s-r)}}\,,
    \label{eq:AntennaFreq}
\end{align}
where $g$, $R$, $r$, and $s$ are the geometric parameters defined in \Cref{fig:SI_MWAntenna}\,(b). 
From \Cref{eq:AntennaFreq}, the resonance frequency increases with the gap $g$, which was adjusted iteratively using a scalpel.
The final $S_{11}$ measurement of the mounted antenna is shown in \Cref{fig:SI_MWAntenna}\,(c) demonstrating alignment with the NV zero-field spin transition at 2.87\,GHz.
For this measurement, the diamond sample pictured in \Cref{fig:SI_MWAntenna}\,(d) was not present on the antenna.

\section{Photon collection efficiency}
\label{app:efficiency}

    Next, we estimate the collection efficiency of NV PL from the dense-NV cavity. Note that these calculations are based on several simplifying assumptions, and there is significant uncertainty in the estimate. We elucidate these assumptions below.
    The NV PL produced inside the dense-NV cavity incurs several sources of loss before it reaches the detector. 
    First, only photons emitted into resonant modes of the cavity are evanescently coupled into the collection fiber-taper\,\cite{Masuda2024}. 
    To estimate that efficiency, we integrate the normalized NV PL spectra collected from both bulk diamond, $I_{\text{bulk}}$, and from the dense-NV cavities, $I_{\text{cav}}$\,(\Cref{fig:2ODMR}\,(a)), with respect to wavelength within the experimentally filtered bounds. 
    Each spectrum is normalized to the maximum intensity in the sampled range. 
    We divide one by the other to determine an upper bound on the spectral collection overlap:
    \begin{align}
        \eta_{\text{spec}} = \frac{\int  I_{\text{cav}} \,\text{d}\lambda}{\int  I_{\text{bulk}}\,\text{d}\lambda} = 6\,\%\,.
    \end{align}
    $\eta_{\text{spec}}$ is likely overestimated by the limited resolution of the spectrometer, which broadens the measured modes.
    Further, although we expect the cavity to induce some preferential emission into the WGMs for PL that spectrally overlaps, much of the PL instead radiates out of the cavity; we do not experimentally measure this radiated emission. 
    Therefore, we treat $\eta_{\text{spec}}$ as an upper bound on the spectral collection efficiency.
    Half of the photons emitted into the WGMs travel clockwise inside the cavity, and the other half travel counter-clockwise.
    Further, only a fraction of the cavity photons are coupled backwards into the fiber---the rest are scattered, lost, or coupled forward. 
    As an upper bound, we estimate that $\eta_{\text{couple}} \approx 50\,\%$ of the produced photons are coupled backwards into the fiber, and the rest are coupled forward into the fiber. 
    This estimate assumes no losses and equal coupling between the forward and backward modes.

Photons can also be scattered or absorbed within the tapered fiber. 
At 780\,nm, the tapered fiber is measured to be $\eta_{\text{fiber}}=40\,(10)\,\%$ transmissive.
We assume that this measured value approximates the transmission efficiency of the tapered fiber across the spectral range of NV PL and that the fiber-taper is equally lossy on either side of where it interacts with the cavity.
In other words, the NV PL intensity at the output of the fiber-taper is scaled by $\sqrt{\eta_{\text{fiber}}}$ by the time it exits the fiber-taper. 
Lastly, there is also loss in the fiber-taper to APD free-space link---we measure the transmission of this link to be $\eta_{\text{free}} = 53\,\%$.
In total, therefore, the collection efficiency is at most
\begin{align}
    \eta_{\text{collect}}  \leq\eta_{\text{spec}}  \,\eta_{\text{couple}} \, \sqrt{\eta_{\text{fiber}}}  \, \eta_{\text{free}} \approx 1.0\,\%.
\end{align}
    
\begin{figure}[t!]
        \centering
        \includegraphics[]{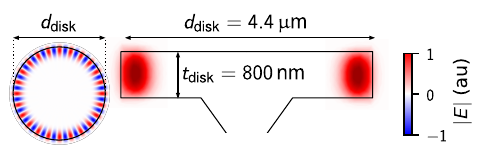}
        \caption{\textbf{Optical mode simulations}. 
        COMSOL Multiphysics simulation of one of the visible optical modes in the dense-NV cavity ($\lambda\sim700\,$nm) predicts an effective mode volume of $V_{\text{eff}} \approx 55\times(\lambda/n_{\text{d}})^3$\,\cite{Mitchell2019} with an azimuthal number of $m=40$, where $n_{\text{d}}$ is the refractive index of diamond. 
        The mode has an effective cross sectional area of $A\sim0.5\,\upmu\text{m}^2$.
        }
        \label{fig:modesimulation}
    \end{figure}

To cross-check this efficiency estimate, we compare the detected PL power with the expected emission rate of NV centers in the cavity.
The detected pulses of NV PL produce a mean amplitude of around 0.10\,V at the highest electronic gain setting (conversion gain of $26.5\times10^6$\,V/W), which corresponds to a detected NV PL optical power of 3.8\,nW, or a detected photon rate of $R_{\text{det}}\approx 1.4\times 10^{10}\,\text{s}^{-1}$.
Further, the NV PL from approximately 1.1 million NV centers couples to the tapered fiber, calculated using the effective mode volume derived from the simulation of one of the visible optical modes shown in \Cref{fig:modesimulation} and the density of NV centers ($7.92\times10^{23}$m$^{-3}$\,\cite{Yu2024}).
Assuming that each NV center near saturation produces $\sim 1$\,million photons every second\,\cite{Hopper2018}, we estimate that the dense-NV cavity produces $R_{\text{prod}} \approx 1.1\times 10^{12}$ photons every second. 
This implies a collection efficiency of 1.3~\%, in reasonable agreement with the 1.0~\% upper bound obtained from the loss estimate described above.
Therefore, around one in 100 NV PL photons is collected and detected, meaning that this platform has great potential to achieve even higher sensitivities by improving the photon collection efficiency through the integration of hybrid photonic waveguides or the other approaches discussed in the main text.

    \begin{figure*}[t!]
    \includegraphics[width=\linewidth]{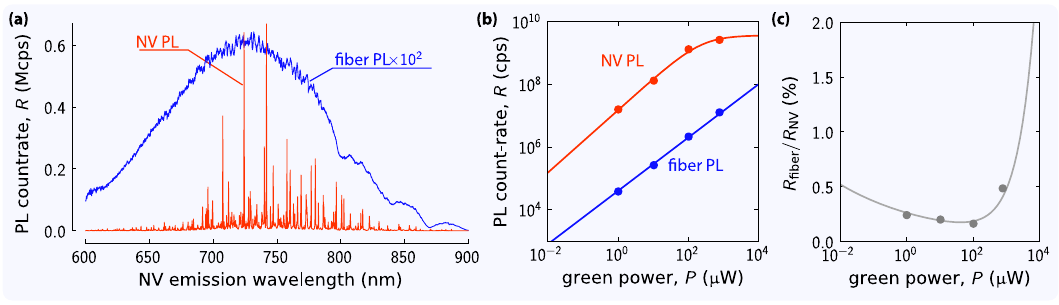}
    \caption{\textbf{Comparison of different sources of the detected photoluminescence. }
    \textbf{(a)} The fiber itself produces photoluminescence in the same frequency range as the NV PL. We plot the photoluminescence spectra both when coupled to the dense-NV cavity (red) and when uncoupled (blue).  
    \textbf{(b)} The photoluminescence from each source---the NV centers and the fiber---is power dependent, and at high powers the fiber PL starts to negatively affect the signal-to-noise ratio of the detected signal. 
    The NV$^{-}$ PL data is fitted using a saturation model: $R \propto P/(P+P_{\text{sat}})$. The fiber PL is fitted using $R\propto P^q$, with fitting parameter $q\sim1$. Here, the PL count-rate $R$ becomes saturated once the green power $P$ exceeds some saturation limit $P_{\text{sat}}$, as demonstrated by the ratio of the two count-rates plotted in \textbf{(c)}.
    \label{fig:SI_PLSpectra}}
    \end{figure*}

\section{Optical saturation of defects and fiber photoluminescence}
\label{app:opticalSaturation}

One important parameter to control when optimizing the sensitivity of a diamond magnetometer is the green optical power used to polarize and read out the state of the NV centers. 
In our system, we both polarize the NV ensemble and collect NV PL using an evanescently coupled fiber-taper.
Designed to be single-mode at telecom wavelengths, the fiber-taper is lossy at visible wavelengths, which reduces our collection efficiency and impedes the sensitivity. 
Moreover, the fiber itself is noisy, as it scatters the pump and produces PL within the frequency range of NV$^{-}$~PL. 
In \Cref{fig:SI_PLSpectra}\,(a), we plot the spectra of the backward-collected light both when coupled to the dense-NV cavity and when uncoupled. 
When uncoupled, the primary source of light is the optical fiber, and, when coupled, the primary source of light is the NV ensemble. 
We plot the pump power dependence of each PL source in \Cref{fig:SI_PLSpectra}\,(b) and the ratio of the two in \Cref{fig:SI_PLSpectra}\,(c).
In the low power regime where the NV centers are unsaturated, NV PL dominates the signal; however, at large green powers the ratio of fiber PL to NV PL begins to increase. 
Effectively, this reduces the contrast of the ODMR, attenuating the SNR, though this effect is small. As shown in \Cref{fig:SI_PLSpectra}~(b), the NV PL count-rate exceeds that of the fiber PL by over two orders of magnitude.

The presence of fiber PL and scattering processes effectively limits how much green power we can put into the cavity, but we mitigate their influence by collecting counter-propagating NV PL and avoiding the collection of collinear Raman scattering.
Firstly, as discussed in \Cref{sec:SI-setup}, we collect counter-propagating PL, filtering it from the pump using a dichroic mirror, a long-pass and a short-pass filter. 
This method improved the SNR and, subsequently, the sensitivity of the magnetometer. 
Further, using an undoped, \textit{pure}, optical fiber could help reduce the PL and improve the sensitivity.

\section{Defect concentration analysis}
\label{app:defectConcentration}
It is possible to model the zero-field splitting of our diamond sample by calculating the eigenenergies of an ensemble of NV centers and substitutional nitrogen defects randomly distributed within a diamond crystal lattice. 
Specifically, we follow the method of Yu et al.\,\cite{Yu2024}, in which we model a volume of diamond, containing a distribution of substitutional nitrogen and NV centers. The electric field at each NV center due to charged defects in the crystal lattice is calculated and is then used to determine the NV center's spin eigenfrequencies. The distribution of the ensemble's eigenfrequencies then determines the zero-field ODMR spectrum and is compared to the measured spectrum. 

To start, our system can be described by the following Hamiltonian for the ground-state spin manifold of the NV center:
\begin{align}
    H = H_{\text{mag}} + H_{\text{nuclear}} + H_{\text{elec|str}},
    \label{eq:hamTotal}
\end{align}
where $H_{\text{mag}}$ describes the electronic spin interaction with a magnetic field, $H_{\text{nuclear}}$ encompasses the hyperfine nuclear spin interaction, and $H_{\text{elec|str}}$ describes the electronic spin interaction with electric fields and crystal lattice strain. 
Respectively, the magnetic, nuclear, and strain components of the Hamiltonian are given by~\cite{Gruber1997,Doherty2013,Barfuss2019,Udvarhelyi2018}
\begin{align}
    \frac{H_{\text{mag}}}{h} 
        &= D S_z^2 + \frac{g_e \mu_B}{h} \,\mathbf{B} \cdot \mathbf{S}, \label{eq:Hmag} \\
    \frac{H_{\text{nuclear}}}{h} 
        &= A_\parallel S_z I_z + A_\perp(S_x I_x + S_y I_y) \notag \\
        &\quad + P_{\text{n}} \left[ I_z^2 - \frac{I(I+1)}{3} \right], \label{eq:Hnuclear} \\
    \frac{H_{\text{elec|str}}}{h} 
        & = (d_\parallel E_z + M_z)S_z^2 \\
        &\quad + (d_\perp E_x + M_x)(S_y^2 - S_x^2) \notag \\
        &\quad + (d_\perp E_y + M_y)(S_x S_y + S_y S_x),
        \label{eq:Helec}
\end{align}
where $D = 2.87$\,GHz is the zero-field ground state splitting parameter, $g_e$ is the NV electronic $g$ factor, $d_\parallel$ and $d_\perp$ are the respective axial and transverse electric dipole moments, $A_\parallel$ and $A_\perp$ are the respective axial and transverse magnetic hyperfine coupling coefficients, $P_{\text{n}}$ is the nuclear electric quadrupole parameter, and $I$ is the nuclear magnetic spin.
Values associated with these terms are summarized in \Cref{table:modelParams}. 
Further, $\mathbf{S} = (S_x, S_y, S_z)$ is the electronic spin-1 operator, $\mathbf{I} = (I_x, I_y, I_z)$ is the nuclear spin operator, and $M_x$, $M_y$, and $M_z$ are spin-strain coupling parameters. 
In \Cref{eq:Helec} we have neglected to write down several other spin-strain coupling parameters which are suppressed by $D$ in our system~\cite{Doherty2013,Barry2020}. 
In the following simulations, for simplicity's sake, we ignore the nuclear hyperfine contribution to the Hamiltonian, $H_{\text{nuclear}}$.

    \begin{table*}[t!]  
    \centering 
    \caption{Parameters and constants used to model the zero field splitting of the NV ensemble.} 
    \begin{tabular}{l l c c} 
    \hline\hline   
    Parameter  & Description & Value & Reference
    \\ 
    \hline\hline  \\ [-0.5ex]   
    $D$  &  zero-field ground state splitting & 2.87\,GHz & \cite{Manson2006}\\
    $g_e$ & NV electronic $g$ factor & 2.003 & \cite{Felton2009}\\
    $\mu_B$ & Bohr magneton & 9.274$\times 10^{-24}$J/T & \cite{Barry2020}\\
    $\gamma/2\pi$  & NV gyromagnetic ratio & 28.0\,GHz/T & \cite{Barry2020}\\
    $d_\parallel$ & axial electric dipole moment & $3.5\times 10^{-3}$\,Hz/(V/m) & \cite{VanOort1990}\\
    $d_\perp$ & transverse electric dipole moment & $0.17$\,Hz/(V/m) & \cite{VanOort1990}\\
    $A_\parallel$ & $^{14}$N  axial magnetic hyperfine coupling coefficient & -2.162\,MHz & \cite{Smeltzer2009}\\
    $A_\perp$ & $^{14}$N  transverse magnetic hyperfine coupling coefficient & -2.70\,MHz & \cite{Felton2009}\\
    $P_{\text{n}}$ & $^{14}$N  nuclear electric quadrupole parameter & -4.945~MHz & \cite{Steiner2010}\\
    $I$ & $^{14}$N  nuclear magnetic spin & 1 & ---\\
    
    [NV] & Nitrogen-vacancy center concentration & 4.5\,ppm & \cite{ElementSix2021}\\
    
    [N$_{\text{s}}^0$] & post-treatment substitutional nitrogen concentration & 8.5\,ppm &\cite{ElementSix2021} \\
    \hline 
    \hline
    \end{tabular} 
    \label{table:modelParams}
    \end{table*}

To solve this Hamiltonian for an ensemble of NV centers, we must determine the electric field local to each NV center, which means we must first detail the distribution of charged defects in such a volume. 
We consider a spherical spatial distribution of substitutional nitrogen and NV centers, according to the nominal concentrations of each ([N$_{\text{s}}^0$] = 8.5\,ppm and [NV] = 4.5\,ppm)~\cite{ElementSix2021}. 
The simulated sphere has a diameter of 200\,nm, containing $7.4\times 10^8$ carbon atoms, 6266 substitutional nitrogen atoms (after processing), and 3317 NV centers. 
We assume that each NV center pairs up with the nearest substitutional nitrogen defect, which provides the NV center with a donor electron and creates a network of charged defect pairs (NV---N$_{\text{s}}$) within the crystal lattice.  
We also assume that the orientations of the NV centers are equally distributed among the four possibilities (see the four crystallographic orientations in \Cref{fig:SI_NV_center}) and that each orientation is randomly spatially distributed. 
Then, we calculate the local electric field, $\mathbf{E}$, at each NV center due to all other charged defects in the crystal lattice. 
To avoid asymmetries at the sphere's surface, we only calculate the fields at NV centers within an 80\,nm sphere, centered within the greater diamond sphere.
Lastly, we assume that each NV center sees a small external magnetic field vector ($|\mathbf{B}| \approx 20$\,$\upmu$T).

\begin{figure}[b!]
    \centering
    \includegraphics[]{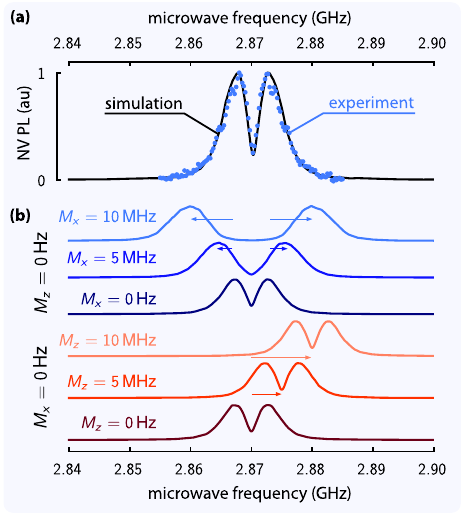}
    \caption{\textbf{Simulating the defect concentration and strain for the dense-NV sensor.} \textbf{(a)} Comparison of the zero-field ODMR spectrum measured from the diamond dense-NV cavity and a simulation of the distribution of spin transition frequencies of NV centers in a diamond sphere, assuming $M_z = 0.5\,$MHz and $|M_x| = 1\,$MHz.
    The electric field at each NV center due to the presence of dipoles is the primary cause of the splitting.
    See the text for more detail. 
    \textbf{(b)} The simulated spectra given other values of axial and transverse strain, assuming $M_x = M_y$.}
    \label{fig:SI_DefectConcentration}
    \end{figure}

Using the calculated electric field, $\mathbf{E}$, local to each NV center in conjunction with the other parameters, we diagonalize \Cref{eq:hamTotal} to solve for the eigenfrequencies of each NV center. To translate the distribution of eigenfrequencies into something we can compare with the measured ODMR spectrum, we convolve each eigenfrequency with a Lorentzian with a linewidth matching that of the Rabi frequency used to perform the measurement ($\Omega_{\text{R}}/2\pi\approx500$\,kHz). 
We then sum each Lorentzian together and normalize to unity. 
This simulation is repeated and averaged over 20 random configurations; a 200~nm sphere keeps each run computationally manageable.
In \Cref{fig:SI_DefectConcentration}\,(a), we compare the measured, normalized zero-field ODMR spectrum of our diamond cavity to that predicted by our simulation. 

    \begin{figure*}[b!]
    \centering
    \includegraphics[width=\linewidth]{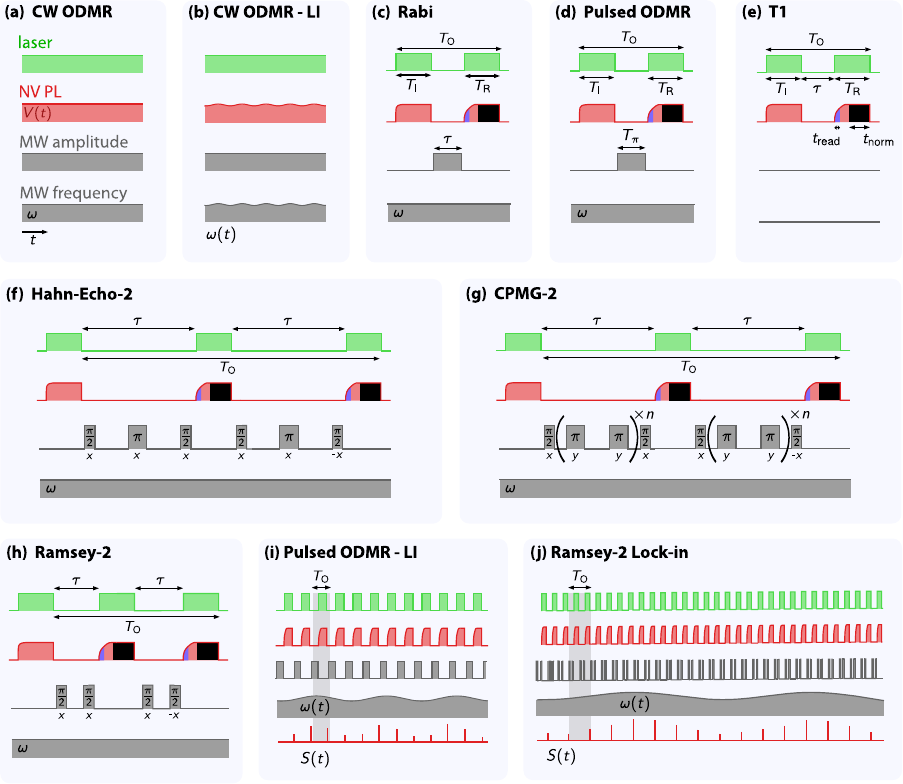}
    \caption{\textbf{The different pulse sequences used to characterize the magnetometer.} For details on each pulse sequence (\textbf{a-j}) see the text in \Cref{app:measSequences}.
    }
    \label{fig:SI_PulseSequences}
    \end{figure*}
Qualitatively, the measured spectrum matches the simulated one well, which achieves a best fit when $M_z = 0.5\,$MHz and $|M_x| = 1\,$MHz.
The simulation produces several insights: firstly, it strengthens our confidence in the values for the density of NV and N$_{\text{s}}$ defects; secondly, it gives us an estimate of the nanofabrication-induced strain.
In \Cref{fig:SI_DefectConcentration}\,(b) we plot the simulated spectra given several other strain parameters, demonstrating not only the effect of different kinds of strain but also the range of strain values that might manifest in the dense-NV cavities. 
Specifically, we see that strain exceeding around 1\,MHz would begin to stretch, translate, or distort the ODMR spectrum into something much different from what we observe, allowing us to establish an upper bound on strain in the dense-NV cavity.
For values of $|M_x| < 1\,$MHz, the magnetic and electric field splitting obscure the effect of strain, forbidding us from ascertaining a more precise value. 
If the strain was higher than $|M_x| \leq 1\,$MHz, the spectrum would be broader than observed.
Given the best fit for the axial strain ($M_z = 0.5\,$MHz), and by assuming that the transverse strain magnitude is equal, we estimate the average total strain on an NV center in the dense-NV sensor to be $M \approx 0.7\,$MHz.
This demonstrates that our fabrication-induced strain is negligible, thereby preserving NV coherence times, which are needed to produce a good quantum sensor.
These results could also assist in future studies focusing on the interactions between defects in materials with high densities of NV centers such as the material studied here.

\section{Measurement sequences}
\label{app:measSequences}

\begin{figure*}[b!]
\centering
\includegraphics[]{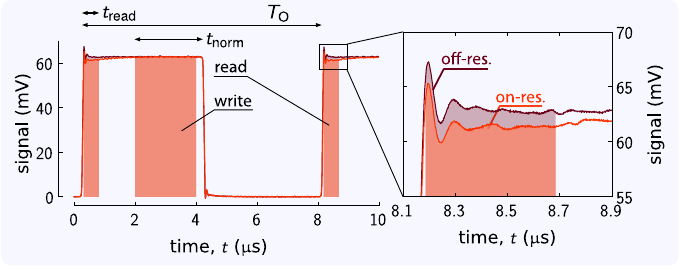}
\caption{Comparison of measured on-resonant and off-resonant ODMR sequences. At left, the red lines plot the amplitude of the NV PL detection during a pulse sequence. We plot two averaged traces---one atop the other---to compare an on-resonant signal and an off-resonant signal. 
To better compare the two signals, we magnify the first few hundred nanoseconds of one pulse, shown on the right. 
The on-resonant microwave pulse (red) results in less NV PL than the off-resonant pulse (dark red). }
\label{fig:SI_Pulses}
\end{figure*}
To demonstrate the versatility of the sensor, we perform several different types of measurements, each of which detects a magnetic-field-dependent change in NV PL.
The duration and sequence of each measurement type are plotted in \Cref{fig:SI_PulseSequences}, where the green laser pulses are plotted in green, followed by the measured NV PL amplitudes in red and the microwave amplitude and frequency in gray. 

\subsection{Continuous-wave measurements}

\def\labelenumi{\alph{enumi})}
\begin{enumerate}
    \item \textbf{Continuous-wave optically detected magnetic resonance (CW-ODMR)}\\
    CW-ODMR is arguably the simplest measurement we perform and has been widely used (see \Cref{fig:SI_PulseSequences}\,(a)). 
    In this sequence, a CW green laser continuously polarizes NV centers into the brighter $|m_s=0\rangle$ ground state, and the microwave frequency, $\omega$, is swept over a range (typically around 2.87\,GHz for systems with low external field).
    Near resonance with one of the $|m_s=0\rangle\rightarrow|m_s=\pm1\rangle$ spin transitions, the microwave drives the NV centers into the darker $|m_s=\pm1\rangle$ states. 
    \item \textbf{CW-ODMR---LI}\\
    The sensitivity of the CW-ODMR measurement can be improved using a lock-in amplifier (see \Cref{fig:SI_PulseSequences}\,(b)). 
    Like before in CW-ODMR, neither the green laser nor the microwave amplitude is pulsed; however, the microwave frequency is time-modulated at frequency $\omega_{\text{LO}}$:
    \begin{equation}
        \omega(\omega_0,t) = \omega_0 + \delta\cos(\omega_{\text{LO}} t).
    \end{equation}
    This modulates the NV PL signal, which is input to a lock-in amplifier alongside a sinusoidal reference signal at frequency $\omega_{\text{LO}}$. 
    The amplitude and frequency of modulation, $\delta$ and $\omega_{\text{LO}}$, respectively, are chosen to maximize the amplitude of the lock-in signal on resonance with a spin transition. 
    By modulating the MW frequency, we suppress low frequency noise and normalize the measurement signal. 
    The modulation frequency for all lock-in measurements ($\sim20\,$kHz) is much higher than the measurement bandwidth ($<1\,$kHz).
    \end{enumerate}

\subsection{Pulsed measurements}
\label{app:pulsedMeas}
A pulsed ODMR sequence typically starts with a green laser pulse, $T_{\text{I}}$ in duration, which is used to polarize the NV ensemble into the $|m_s=0\rangle$ state.
Polarization is followed by a series of microwave pulses, used to control the electronic spin state. Finally, a second green pulse, $T_{\text{R}}$ in duration, reads out the spin state of the ensemble.
Experimentally, this manifests in a reduction in the detected voltage signal that lasts until the ensemble re-polarizes into the $|m_s=0\rangle$ state. 
For a sufficiently intense pump field, the reduction in signal will last for approximately 300\,ns.
The length of reduction is determined by the room temperature lifetime of the singlet state and the green pump intensity, which determines how quickly the NV center cycles.
We integrate the signal over this duration ($t_{
\text{read}}$), shaded blue in \Cref{fig:SI_PulseSequences} and labeled in \Cref{fig:SI_PulseSequences}\,(e). 
Further, because of intracavity pump power fluctuations, we normalize the signal to the end of the pulse, shaded in black in \Cref{fig:SI_PulseSequences}. 
Experimentally, we find $t_{\text{read}}=500$\,ns, $t_{\text{norm}}=2$\,$\upmu$s, and $T_{\text{I}}=T_{\text{R}}=4$\,$\upmu$s to be optimal for all pulsed sequences---for more discussion of optimization, we point the reader to the review article by Barry et al.\,\cite{Barry2020}. 
The ODMR signal for any single pulsed measurement is given by
\begin{align}
    S = \frac{\int_{t_{\text{read}}}V(t)\,dt}{\int_{t_{\text{norm}}}V(t)\,dt}.
\end{align}
As an example, we plot a pulsed measurement in \Cref{fig:SI_Pulses}, where a resonant microwave pulse is compared to a non-resonant microwave pulse. 
In the optimized measurements, as in \Cref{fig:SI_Pulses}, the first polarizing pulse is removed to reduce the overhead time of the pulse sequence. 
Instead, we assume that the previous pulse fully re-polarizes the ensemble.

    \begin{enumerate}[resume]
    \item \textbf{Rabi}\\
    A Rabi sequence is used to characterize the microwave pulse length required to invert the spin population, $T_\pi$. 
    A laser pulse is followed by a microwave pulse of duration $\tau$. 
    The spin state is then read-out using another laser pulse. 
    To observe Rabi flopping, $\tau$ is step-wise tuned, and the signal contrast for each $\tau$ is measured. 
    An example of this measurement is shown in the inset in \Cref{fig:2ODMR}\,(b) in the main text. 
    The microwave frequency $\omega$ is fixed on resonance with the spin transition throughout.
    \item\textbf{Pulsed ODMR}\\
    Pulsed ODMR measures the spin transition frequency, as was done using CW-ODMR; however, pulsed ODMR can achieve better contrast and  sensitivity\,\cite{Barry2020}. 
    Here, the pulse sequence is the same as for Rabi, except that instead of sweeping the microwave pulse duration $\tau$, the microwave frequency $\omega$ is swept, and the duration of the microwave pulse is kept constant ($T_{\pi}$). 
    The microwave power and duration are chosen such that $T_{\pi}\approx T_2^*$ ($\Omega_{\text{R}}/2\pi \approx 1$\,MHz).
    \item{$\mathbf{T_1}$}\\
    The longitudinal relaxation time, $T_1$, is characterized using a pulse sequence similar to that used in pulsed ODMR, however, no microwave field is applied.
    A green pulse initializes the spin population into the $|0\rangle$ state, and is followed by a free-precession period, $\tau$. 
    After that time, the spin population is read-out using another green pulse. 
    By step-wise tuning $\tau$, we determine $T_1$.
    \item\textbf{Hahn-echo-2}\\
    We use a two-pulse Hahn-echo sequence to measure the transverse relaxation or decoherence time ($T_2$) of the NV spin ensemble. 
    The entire sequence consists of a polarization pulse, a series of microwave pulses, a read-out/polarization pulse, a second series of microwave pulses, and a final read-out pulse. 
    Each series of time-symmetric microwave pulses consists of a $\pi/2$-pulse ($x$) followed by a $\pi$-pulse ($x$) and a second $\pi/2$-pulse ($x$). 
    The second pulse series differs only in the phase of the final $\pi/2$ pulse ($-x$). 
    By inverting the phase and subtracting the signal from the second pulse sequence from that of the first, we suppress noise. 
    \item\textbf{CPMG-2}\\
    We use a two-pulse CPMG sequence to decouple the spins from the magnetic environment, thereby extending the coherence time\,\cite{Bar-Gill2013}. 
    Each series of time-symmetric microwave pulses consists of a $\pi/2$-pulse ($x$) followed by $2n$ $\pi$-pulses ($y$) and a second $\pi/2$-pulse ($x$). 
    Each $\pi$-pulse ($y$) is separated in time by approximately $\tau/2n$, and the first and last $\pi$-pulses ($y$) are separated from the $\pi/2$-pulses ($x$) in time by approximately $\tau/4n$.
    The duration $\tau$ includes not only the time delays but also the duration of all microwave pulses.
    The second pulse series differs only in the phase of the final $\pi/2$-pulse ($-x$), which is used to suppress common-mode noise.
    \item\textbf{Ramsey-2} \\
    Our Ramsey-2 sequence is identical to the Hahn-echo-2 sequence, excluding the refocusing $\pi$ pulses. 
    Ramsey sequences use two $\pi/2$-pulses separated by phase acquisition time, $T_{\varphi}$, to measure spin contrast. Either $T_{\varphi}$ or $\omega$ can be swept, depending on the measurement. 
    Similar to Hahn-echo-2 and CPMG-2, a second pulse sequence with an inverted final pulse is used to suppress noise.
    \item\textbf{Pulsed ODMR-LI} \\
    Similar to a CW measurement, a lock-in measurement can be used to suppress noise in a pulsed ODMR scheme\,\cite{Zhang2022}. 
    To do so, we perform several successive pulse sequences [in \Cref{fig:SI_PulseSequences}\,(i), a single pulse scheme is highlighted by a translucent grey vertical rectangle], and, at the same time, we discretely modulate the microwave frequency. 
    Over a single pulse sequence, the microwave frequency is constant, but each subsequent measurement uses a slightly different microwave frequency to produce a modulated pulsed ODMR signal.
    For this to work, the modulation frequency must be significantly lower than the pulsed ODMR sampling frequency, as determined by the Nyquist bound.
    The discrete measurement series produced in time, shown in \Cref{fig:SI_PulseSequences}\,(i) by a discrete series of vertical red bars, is input to a lock-in amplifier, which is referenced to the modulation frequency.
    \item\textbf{Ramsey-2 Lock-in} \\
    As was done with the pulsed ODMR-LI scheme, we can also lock-in a Ramsey scheme, effectively replacing a single pulsed sequence with a Ramsey-2 sequence. 
    In \Cref{fig:SI_PulseSequences}\,(j), a single Ramsey-2 pulse scheme is highlighted by a translucent grey vertical rectangle. 
\end{enumerate}

\section{Magnetic sensitivity and optimization}
\label{app:sensitivityOptimization}

 \begin{figure*}[t!]
    \centering
    \includegraphics[width = \linewidth]{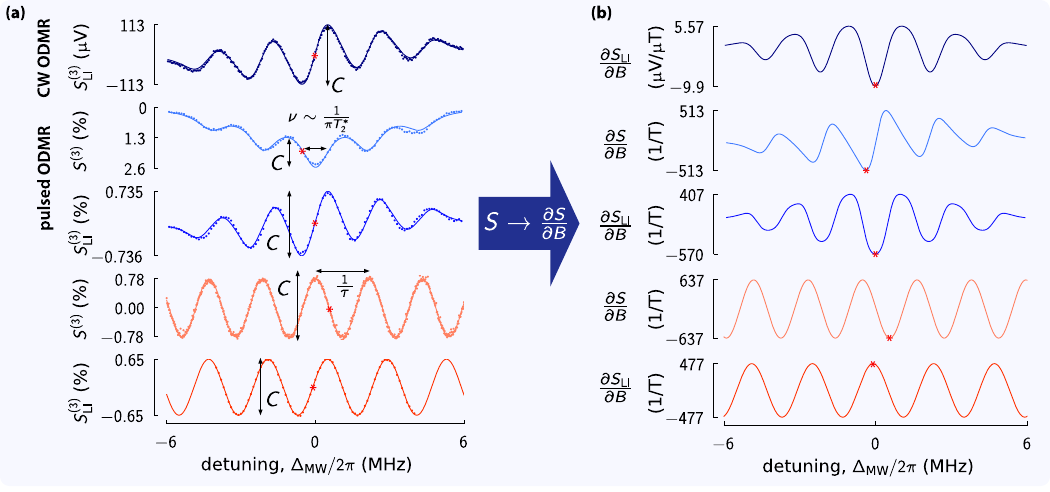}
    \caption{Estimating magnetic sensitivity. 
    \textbf{(a)} For each protocol we perform ODMR measurements as a function of microwave detuning, $\Delta_{\text{MW}}/2\pi$, allowing us to determine the signal response $\partial S$ to a small change in the axial magnetic field $\partial B$. 
    For each measurement protocol we identify the position of maximum response using a red asterisk. 
    Fits using the models discussed in the appendix of each signal give the various parameters used to determine the photon-shot-noise-limited sensitivities. \textbf{(b)} The maximum signal responses to a small perturbation in magnetic field are determined from the fits in (a). The detuning is converted to a magnetic-field-dependence using $B ={\Delta_{\text{MW}}}/\gamma$.
    }
    \label{fig:SI_sensitivityExtraction}
\end{figure*}
Generally, the sensitivity of a magnetic field measurement is given by\,\cite{Budker2007,Taylor2008}
\begin{align}
    \eta = \frac{\sigma(t) \sqrt{t}}{\frac{\partial S}{\partial B}},
\end{align}
where $\sigma$ is the standard deviation of a measurement signal $S$, whose response $\partial S$ depends on a small change in the axial magnetic field $\partial B$, and $t$ is the total measurement time
of the magnetic field. 
The calculation of this sensitivity greatly depends on the experimental implementation and especially on the type of magnetic field measurement. 
In CW-ODMR, for example, the photon-shot-noise-limited sensitivity is given by\,\cite{Drau2011,Barry2016}
\begin{align}
    \eta_{\text{cw}} \approx \frac{8\pi}{3\sqrt{3}}\frac{1}{\gamma}\frac{\nu}{ C\sqrt{R}} ,
    \label{eq:CWSensitivity}
\end{align}
where $\nu$ is the ODMR linewidth, $C$ is the ODMR contrast, $R$ is the photon detection rate, $\gamma/2\pi = g_e\mu_B/h$ is NV gyromagnetic ratio, $h$ is Planck's constant, $g_e$ is the NV$^{-}$ electronic $g$ factor, and $\mu_B$ is the Bohr magneton. 
Values for these constants are presented in \Cref{table:modelParams}.
Two other DC magnetometry sequences are used in this manuscript: a pulsed ODMR sequence, and a Ramsey ODMR sequence. For a pulsed ODMR sequence, the photon-shot-noise-limited sensitivity is given by\,\cite{Barry2016}
\begin{align}
    \eta_{\text{pulsed}} \approx \frac{8\pi}{3\sqrt{3}} \frac{1}{\gamma} \frac{1}{C\sqrt{\mathcal{N}}}\frac{\sqrt{T_{\text{O}}}}{\pi T_2^*},
    \label{eq:PulsedSensitivity}
\end{align}
and for a Ramsey sequence, the photon-shot-noise-limited sensitivity is given by\,\cite{Barry2016}
\begin{align}
    \eta_{\text{ramsey}} \approx \frac{1}{\gamma}\frac{1}{C\sqrt{\mathcal{N}}}\frac{\sqrt{T_{\text{O}}}}{\tau}.
    \label{eq:RamseySensitivity}
\end{align}
In \Cref{eq:RamseySensitivity,eq:PulsedSensitivity}, $T_{\text{O}}$ is the sequence overhead time, and $\mathcal{N}=Rt_{\text{read}}$ is the number of detected photons per measurement. 
The phase integration time, $\tau$, in the case of the Ramsey scheme, is set to 370\,ns\,$\sim T_2^*$, which corresponds to a point of optimal signal contrast. 

The parameters required to determine the shot-noise-limited sensitivities given by \Cref{eq:CWSensitivity,eq:RamseySensitivity,eq:PulsedSensitivity} can be determined by fitting the ODMR curves of the respective measurement protocols. 
We plot detuning measurements for each protocol investigated in this manuscript in \Cref{fig:SI_sensitivityExtraction}\,(a) and define the respective measurement parameters for each protocol, which can be determined by curve-fitting the data.
The models used to fit the ODMR spectra depend on the type of measurement performed.
For example, ODMR data acquired while driving at a single microwave frequency, such as that shown in \Cref{fig:sens} can be modeled by
\begin{align}
\begin{split}
    S^{(1)}(\Delta_{\text{MW}}, C, \nu) = C[&\mathcal{L}(\Delta_{\text{MW}}-\Omega_{\text{HF}}, \nu) \\
\quad&+ \mathcal{L}(\Delta_{\text{MW}}, \nu) \\
\quad& + \mathcal{L}(\Delta_{\text{MW}} + \Omega_{\text{HF}}, \nu)]\,,
    \end{split}
    \label{eq:fit_ODMR_1}
\end{align}
where $C$ is the signal amplitude and 
\begin{align}
    \mathcal{L}(\Delta, \nu) \coloneqq \frac{\left(\frac{\nu}{2}\right)^2}{\Delta^2 + \left(\frac{\nu}{2}\right)^2}\label{eq:fit_lorentzian}
\end{align}
is a Lorentzian centered at zero with a full-width at half-maximum (FWHM) equal to the hyperfine linewidth, $\nu$.
Further, ODMR spectra produced while simultaneously driving all three hyperfine transitions are fit using
\begin{equation}
\begin{split}
    S^{(3)}(\Delta_{\text{MW}}, C, \nu) = \frac{C}{3} \Big[ &\mathcal{L}(\Delta_{\text{MW}}-2\Omega_{\text{HF}}, \nu)  \\
    \quad&+2\mathcal{L}(\Delta_{\text{MW}}-\Omega_{\text{HF}}, \nu) \\
    \quad&+3\mathcal{L}(\Delta_{\text{MW}}, \nu) \\   
    \quad&+2\mathcal{L}(\Delta_{\text{MW}}  + \Omega_{\text{HF}}, \nu) \\
    \quad&+\mathcal{L}(\Delta_{\text{MW}} + 2 \Omega_{\text{HF}}, \nu)\Big]\,.
\end{split}\label{eq:fit_ODMR_3}
\end{equation}
    \setlength{\extrarowheight}{2pt}
    \begin{table*}[t!]
    \centering
    \caption{Summary of fitted parameters for all measurement protocols.
    }
    \begin{tabular}{l|cccccccc}
    \hline \hline
    protocol  & $C$ (\%)  & $\nu$ (MHz) & $R$ (ns$^{-1}$) & $T_2^*$ ($\upmu$s) & $t_{\text{read}}$ ($\upmu$s) & $T_{\text{O}}$ ($\upmu$s) & $\tau$ ($\upmu$s) & $\eta$ (nT/$\sqrt{\text{Hz}}$)  \\ \hline \hline
    CW     & 0.25\,(1) & 1.65\,(3)   & 7\,(1)          & ---                 & ---                           & ---                        & ---                & 220\,(40)  \\
    pulsed    & 1.25\,(4)  & 1.45\,(2)  & 8\,(1)             & 0.43\,(2)                 & 0.5                           & 8                        & ---                &    ---     \\
    pulsed-LI & 1.47\,(5)  & 1.25\,(3)  & 8\,(1)          & 0.43\,(2)                 & 0.5                          & 8                         & ---                &        71\,(10)      \\
    Ramsey    & 1.56\,(5)  & ---          & 8\,(1)          & 0.43\,(2)          & 0.5                          & 8                         & 0.37              &      ---              \\
    Ramsey-LI & 1.30\,(5)  & ---          & 8\,(1)          & 0.43\,(2)          & 0.5                          & 8                         & 0.37              & 52\,(8) 
    \label{table:fittedParameters}
    \end{tabular}
    \end{table*}
    
    \begin{figure*}[b!]
    \centering
    \includegraphics[width = \textwidth]{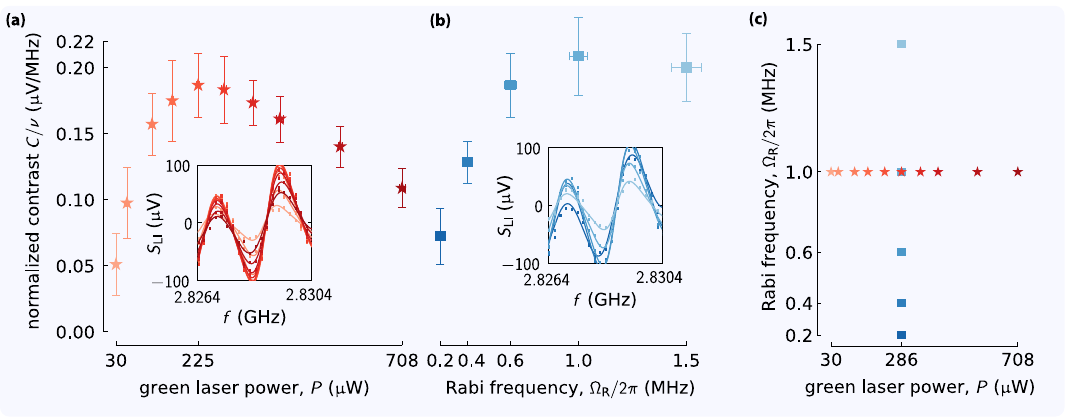}
    \caption{\textbf{Optimization of CW-ODMR.} To achieve optimized ODMR, the ratio of contrast and linewidth, $C/\nu$, is maximized as a function of both green laser power \textbf{(a)} and microwave power, presented as Rabi frequency \textbf{(b)}. The contrast and linewidth are both extracted from the CW-Lock-in ODMR spectra characterized at each green laser power and Rabi frequency. The ODMR spectra used for characterization are fitted and plotted in the insets shown in each plot. 
    The colors of the scatter points and fits in the insets correspond with the scatter points of the same color in the primary figure, denoting the powers used in each measurement. 
    \textbf{(c)} The sampled parameters for the data in \textbf{(a)} and \textbf{(b)}---while the green power is swept at the optimal Rabi frequency, the Rabi frequency is swept using a non-optimal green laser power (286\,$\upmu$W instead of 225\,$\upmu$W).
    }
    \label{fig:SI_CWOptimization}
    \end{figure*}
The pulsed ODMR data shown in \Cref{fig:SI_sensitivityExtraction}\,(a) was fit using \Cref{eq:fit_ODMR_3}.
Following from \Cref{eq:fit_ODMR_3}, the equations used to fit the lock-in pulsed and CW-ODMR spectra are given by
\begin{equation}
\begin{split}
    S^{(3)}_{\text{LI}}(\Delta_{\text{MW}}, C, \nu, \delta) = &S^{(3)}(\Delta_{\text{MW}}-\delta, C, \nu)  \\
    \quad&- S^{(3)}(\Delta_{\text{MW}}+\delta, C, \nu)\,,\end{split}\label{eq:fit_ODMR_3_LI}
\end{equation}
where $\delta \sim 0.5$\,MHz is the amplitude of the microwave frequency modulation. Equation\,(\ref{eq:fit_ODMR_3_LI}) is used to fit the lock-in pulsed and CW-ODMR spectral data shown in \Cref{fig:SI_sensitivityExtraction}\,(a).
Lastly, the Ramsey ODMR spectra are fitted using the following equation:
\begin{align}
    S^{(3,\text{Ramsey})}(\Delta_{\text{MW}}, C,\tau) = C\cos\left(\tau \Delta_{\text{MW}} + \theta \right)\,,\label{eq:fit_ODMR_Ramsey}
\end{align}
where, as before, $\tau$ is the time delay or phase integration time of the Ramsey protocol, and $\theta$ is a phase offset. 
The Ramsey data in \Cref{fig:SI_sensitivityExtraction}\,(a) is fitted using \Cref{eq:fit_ODMR_Ramsey}, and its lock-in counterpart data is fitted using 
\begin{equation}
\begin{split}
    S^{(3,\text{Ramsey})}_{\text{LI}}(\Delta_{\text{MW}}, C, \tau, \delta) = &S^{(3,\text{Ramsey})}(\Delta_{\text{MW}}-\delta, C, \tau) \\
    \quad&- S^{(3,\text{Ramsey})}(\Delta_{\text{MW}}+\delta, C, \tau)\,.\end{split}\label{eq:fit_ODMR_Ramsey_LI}
\end{equation}
The parameters extracted by fitting \Cref{eq:fit_ODMR_3,eq:fit_ODMR_3_LI,eq:fit_ODMR_Ramsey,eq:fit_ODMR_Ramsey_LI} are then used in conjunction with \Cref{eq:CWSensitivity,eq:RamseySensitivity,eq:PulsedSensitivity} to predict the shot-noise-limited sensitivities---we present these parameters and calculated values in \Cref{table:fittedParameters}. 

The detuning curves in \Cref{fig:SI_sensitivityExtraction}\,(a) can also be used to extract the maximum signal response to a magnetic field perturbation $(\partial S/\partial B)$. 
We plot the scaled derivatives of the fits in \Cref{fig:SI_sensitivityExtraction}\,(b) and indicate the positions of maximum sensitivity using red asterisks. 
These fits are used to convert the ODMR signals into magnetic field signals shown in \Cref{fig:sens}\,(d, e) in the primary manuscript. 
Detuning is converted into a magnetic field perturbation by dividing it by the NV gyromagnetic ratio, $\gamma$.

We extract the inhomogeneous dephasing time, $T_2^*$, from the Ramsey measurements shown in \Cref{fig:char}\,(a) by fitting the data to the following equation:
\begin{align}
\begin{aligned}
S(\tau) = e^{-\tau/T_2^*} \Big\{
&A_1 \cos\big[(\Delta_{\text{MW}}+\Omega_{\text{HF}}) \tau + \phi_1\big] \\
& + A_2 \cos\big[(\Delta_{\text{MW}}-\Omega_{\text{HF}}) \tau + \phi_2\big] \\
& + A_3 \cos\big[\Delta_{\text{MW}} \tau + \phi_3\big]\Big\}.
\end{aligned}
\end{align}
    
\begin{figure*}[t!]
\centering
\includegraphics[width=\linewidth]{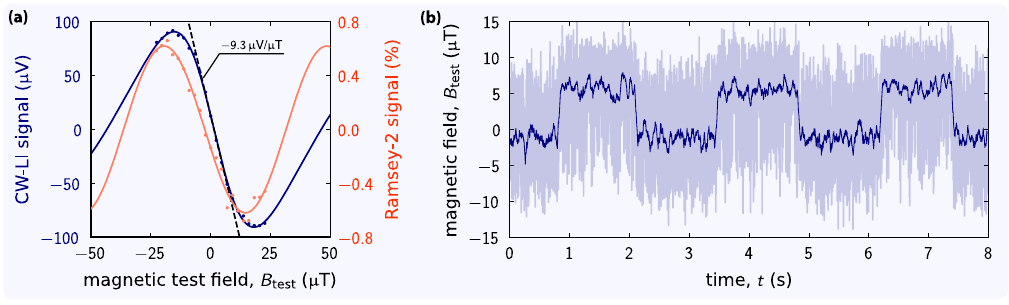}
\caption{\textbf{Measurement of a magnetic test field.} \textbf{(a)} ODMR signal produced by both the CW and Ramsey protocols as a function of applied magnetic test field from a nearby solenoid. \textbf{(b)} Magnetic field signal measured using a CW protocol as a magnetic test field is switched on and off. 
The light blue trace shows the signal measured at 1\,ms intervals, whereas the dark blue trace shows a  highly averaged signal measured at 50\,ms intervals. }
\label{fig:SI_solenoid}
\end{figure*}
\subsection{Optimization}
\label{app:optimization}

Optimization of the CW protocol [\Cref{eq:CWSensitivity}] involves maximizing the count-rate, $R$, as well as the ratio of contrast and linewidth, $C/\nu$.
Both of these parameters are dependent on the optical control field intensity, as well as the microwave control power, both of which we tune to optimize the sensitivity. 
The results shown in \Cref{fig:SI_CWOptimization} were used to optimize the experiment: we found that the optimal experiment used  CW green power $P=225\,\upmu$W input to the fiber-taper and microwave power that produced a Rabi frequency $\Omega_{\text{R}}/2\pi=1.0\,$MHz.
At 532~nm we measure the fiber-transmission to be roughly equal to that measured at 780~nm, so we estimate the power that reaches the tapered region to be $P_{\text{disk}}\sim 142\,(36)\,\upmu$W. 
As discussed in~\Cref{app:cavityFiberCoupling}, coupling to the fiber is very weak and most of the light transmits directly through the fiber, however, we cannot estimate the power scattered into the cavity. 
By engineering a cavity mode at the laser wavelength and using a waveguide tailored to carry both the pump and NV PL, the required optical powers would be greatly reduced.
The optimized parameters used for the CW method in conjunction with \Cref{eq:CWSensitivity} predict a sensitivity of $220\,(40)\,$nT/$\sqrt{\text{Hz}}$.
We perform similar optimizations for the pulsed and Ramsey protocols.
Their optimized experimental parameters in conjunction with \Cref{eq:PulsedSensitivity,eq:RamseySensitivity} predict respective shot-noise-limited sensitivities of $71\,(10)\,$nT/$\sqrt{\text{Hz}}$ and $52\,(8)\,$nT/$\sqrt{\text{Hz}}$, which are within uncertainty of the measured values.

\section{Measurement of a magnetic test field}
\label{app:testField}

To demonstrate the use of the dense-NV cavity as a magnetometer, we use it to measure a test magnetic field produced by a solenoid placed $\sim 10\,$cm from the diamond sample.
We sweep the current passing through the solenoid and measure the resulting ODMR signal using both the CW and Ramsey protocols. 
The current-dependent magnetic field produced by the solenoid is calibrated using a gaussmeter (Hirst GM09), allowing us to produce the calibration curves plotted in \Cref{fig:SI_solenoid}\,(a). 
We then measure a time-varying magnetic field by switching the solenoid on and off and measuring the response using the CW protocol, as shown in \Cref{fig:SI_solenoid}\,(b). 
This concretely demonstrates that the dense-NV cavity sensor is capable of detecting weak external magnetic fields.

\bibliography{bibliographies/main}

\end{document}